\documentclass[journal]{IEEEtran}

\usepackage{graphicx} 
\usepackage{epstopdf}
\usepackage{bbm}
\usepackage{cuted}
\usepackage{gensymb}
\include{macros}
\usepackage{amsmath,amssymb,epsfig,psfrag,cite,subfigure}

\usepackage{xr}
\usepackage{amsmath,amssymb,color}
\usepackage{multirow}
\usepackage{mathtools}
\usepackage{enumerate}

\usepackage{enumitem}

\usepackage{lettrine}

\usepackage{float} 
\usepackage{wrapfig} 

\usepackage{algorithm}
\usepackage{algpseudocode}

\usepackage{bm}
\usepackage[multiple]{footmisc}

\usepackage{mathtools}

\newtheorem{proof}{Proof}

\newenvironment{axioms}
{\enumerate[label=\textit{A\arabic*.}, ref=A\arabic*]}
{\endenumerate}
\makeatletter
\newcommand\varitem[1]{\item[\textbf{A\arabic{enumi}\rlap{$#1$}.}]%
	\edef\@currentlabel{A\arabic{enumi}{$#1$}}}
\makeatother

\newcommand{\y}{\mathbf{y}}

\newcommand{\ux}{\mathbf{x}}
\newcommand{\uy}{\mathbf{y}}
\newcommand{\ua}{\mathbf{a}}
\newcommand{\ub}{\mathbf{b}}

\newcommand{\expectation}{{\mathbb{E}}}

\newcommand{\JJJ}{\mathbf{J}}

\newcommand{\un}{\mathbf{n}}
\newcommand{\uni}{\mathbf{n}^{(i)}}
\newcommand{\tilden}{\widetilde{\mathbf{n}}}
\newcommand{\tildenTj}{\widetilde{\mathbf{n}}_{T,j}}

\newcommand{\Skij}{S_k^{(i,j)}}
\newcommand{\tildeSkj}{\widetilde{S}_k^{(j)}}

\newcommand{\tildePlkj}{\widetilde{P}_{l,k}^{(j)}}
\newcommand{\Plkij}{P_{l,k}^{(i,j)}}
\newcommand{\tildePTl}{\widetilde{P}_{T,l}}
\newcommand{\PTli}{{P}_{T,l}^{(i)}}
\newcommand{\tildephilkj}{\widetilde{\phi}_{l,k}^{(j)}}
\newcommand{\tildephilkij}{{\phi}_{l,k}^{(i,j)}}
\newcommand{\tildethetalkj}{\widetilde{\theta}_{l,k}^{(j)}}
\newcommand{\tildethetalkij}{{\theta}_{l,k}^{(i,j)}}
\newcommand{\tildenlkj}{\widetilde{\eta}_{l,k}^{(j)}}
\newcommand{\nlkij}{\eta_{l,k}^{(i,j)}}
\newcommand{\nRkj}{{\mathbf{n}}_{R,k}^{(j)}}
\newcommand{\nTli}{{\mathbf{n}}_{T,l}^{(i)}}

\newcommand{\utildedlkj}{{\widetilde{\mathbf{d}}}_{l,k}^{(j)}}
\newcommand{\udlkij}{\mathbf{d}_{l,k}^{(i,j)}}

\newcommand{\Akj}{A_k^{(j)}}

\newcommand{\tildeml}{\widetilde{m}_{l}}
\newcommand{\mli}{m_l^{(i)}}

\newcommand{\tildealphalkj}{\widetilde{\alpha}_{l,k}^{(j)}}
\newcommand{\alphalkij}{{\alpha}_{l,k}^{(i,j)}}

\newcommand{\Skijgen}[3]{S_{#1}^{(#2,#3)}}
\newcommand{\tildeSkjgen}[2]{\widetilde{S}_{#1}^{(#2)}}

\newcommand{\norm}[1]{\big\lVert#1\big\rVert}
\newcommand{\normbigg}[1]{\bigg\lVert#1\bigg\rVert}

\newcommand{\Ntotjk}{N_{tot}^{(j,k)}}
\DeclareMathOperator*{\argmin}{arg\,min}

\newcommand{\tildenTjgen}[1]{\widetilde{\mathbf{n}}_{T,#1}}

\newcommand{\unTgen}[2]{\mathbf{n}^{(#1)}_{T,#2}}

\newcommand{\nRkjgen}[2]{\mathbf{n}_{R,#1}^{(#2)}}

\newcommand{\tildemlgen}[1]{\widetilde{m}_{#1}}

\newcommand{\mligen}[2]{m_{#1}^{(#2)}}

\newcommand{\xx}{{\mathbf{x}}}
\newcommand{\yy}{{\mathbf{y}}}
\newcommand{\zz}{{\mathbf{z}}}
\newcommand{\ww}{{\mathbf{w}}}

\newcommand{\nr}{{\mathbf{n}_R}}
\newcommand{\nt}{{\mathbf{n}_T}}
\newcommand{\xtheta}{{ \bm{\theta} }}

\newcommand{\mtA}{{\mathcal{L}}}
\newcommand{\mtB}{{ \mathcal{B}_{\alpha} }}
\newcommand{\mtN}{{\mathcal{N}}}
\newcommand{\mtC}{{\mathcal{C}}}

\newcommand{\mtF}{{\mathcal{F}}}
\newcommand{\mtG}{{\mathcal{G}}}
\newcommand{\mtS}{{\mathcal{S}}}
\newcommand{\mtFtilde}{{\mathcal{\widetilde{F}}}}
\newcommand{\mtFbreve}{{\mathcal{\breve{F}}}}

\newcommand{\setfromone}[1]{{{\{1,2,\ldots,#1\}}}}

\newcommand{\gammatildelkj}{{\tilde{\gamma}_{\ell,k}^{(j)} }}

\newcommand{\gammalkij}{{\gamma_{\ell,k}^{(i,j)} }}

\newcommand{\xxit}[2]{{ \xx_{#1}^{(#2)} }}

\newcommand{\glkjfunc}{{ \tilde{g}_{\ell,k}^{(j)} }}
\newcommand{\glkijfunc}{{ g_{\ell,k}^{(i,j)} }}
\newcommand{\Gflambda}[2]{{ G_{#1}^{#2} }}
\newcommand{\Hflambda}[1]{{ H_{#1} }}

\newcommand{\tildekappa}{{ \tilde{\kappa}_{\ell,k}^{(j)} }}
\newcommand{\kappalkij}{{ \kappa_{\ell,k}^{(i,j)} }}

\newcommand{\tildexjn}{{ \tilde{\xx}_{j}^{(n)} }}
\newcommand{\lambdajn}{{ \lambda_{j}^{(n)} }}

\newcommand{\xthetajn}{{ \xtheta_j^{(n)} }}

\newcommand{\tildexjnzeta}{{ \tilde{\xx}_{j}^{(n(\zeta))} }}
\newcommand{\fcirc}{{ f^{\circ} }}
\newcommand{\tildezeta}{{ \tilde{\zeta} }}

\newcommand{\tildexjnt}{{ \tilde{\xx}_{j}^{(n_t)} }}

\graphicspath{{./Figures/}}

\begin{document}

\renewcommand{\thepage}{}


\title{Cooperative Localization in Visible Light Networks: Theoretical Limits and Distributed Algorithms}

\author{Musa Furkan Keskin, Osman Erdem,\thanks{The authors are with the Department of Electrical and Electronics Engineering, Bilkent University, 06800, Ankara, Turkey, Tel: +90-312-290-3139, Fax: +90-312-266-4192, Emails: \{keskin,oerdem,gezici\}@ee.bilkent.edu.tr.}
and 
Sinan Gezici, \textit{Senior Member, IEEE}\thanks{Part of this work was presented at IEEE International Black Sea Conference on Communications and Networking (BlackSeaCom), June 2017, www.ee.bilkent.edu.tr/$\sim$gezici/coopVLP.pdf}\vspace{-0.7cm}}


\maketitle

\begin{abstract}
	Light emitting diode (LED) based visible light positioning (VLP) networks can provide accurate location information in indoor environments. In this manuscript, we propose to employ cooperative localization for visible light networks by designing a VLP system configuration that involves multiple LED transmitters with known locations (e.g., on the ceiling) and visible light communication (VLC) units equipped with both LEDs and photodetectors (PDs) for the purpose of cooperation. First, we derive the Cram\'er-Rao lower bound (CRLB) and the maximum likelihood estimator (MLE) for the localization of VLC units in the proposed cooperative scenario. To tackle the nonconvex structure of the MLE, we adopt a set-theoretic approach by formulating the problem of cooperative localization as a quasiconvex feasibility problem, where the aim is to find a point inside the intersection of convex constraint sets constructed as the sublevel sets of quasiconvex functions resulting from the Lambertian formula. Next, we devise two feasibility-seeking algorithms based on iterative gradient projections to solve the feasibility problem. Both algorithms are amenable to distributed implementation, thereby avoiding high-complexity centralized approaches. Capitalizing on the concept of quasi-Fej\'er convergent sequences, we carry out a formal convergence analysis to prove that the proposed algorithms converge to a solution of the feasibility problem in the consistent case. Numerical examples illustrate the improvements in localization performance achieved via cooperation among VLC units and evidence the convergence of the proposed algorithms to true VLC unit locations in both the consistent and inconsistent cases.
	
	\textbf{\textit{Index Terms}--} Positioning, visible light, cooperative localization, set-theoretic estimation, quasiconvex feasibility, gradient projections, quasi-Fej\'er convergence.
\end{abstract}

\vspace{-0.4cm}

\section{Introduction}\label{sec:intro}


\subsection{Background and Motivation}
Accurate wireless positioning plays a decisive role in various location-aware applications, including patient monitoring, inventory tracking, robotic control, and intelligent transport systems \cite{WSNloc_book,pahlavan2013principles,HLIU,Sinan_Survey}. In the last two decades, radio frequency (RF) based techniques have commonly been employed for wireless indoor positioning \cite{HandbookPos,bookSahin}. Recently, light emitting diode (LED) based visible light positioning (VLP) networks have emerged as an appealing alternative to RF-based solutions, providing high-accuracy and low-cost indoor localization services \cite{VLP_Roadmap}. While visible light networks can be harnessed for enhancing localization performance in indoor scenarios \cite{SurveyVLC15}, they also offer illumination and high speed data communications simultaneously without incurring additional installation costs via the use of existing LED infrastructure \cite{BeyondPoint}. In indoor VLP networks, various position-dependent parameters such as received signal strength (RSS) \cite{VLP_CRLB_RSS,zhang2014asynchronous}, time-of-arrival (TOA) \cite{CRB_TOA_VLC,MFK_CRLB}, time-difference-of-arrival (TDOA) \cite{TDOA_VLC} and angle-of-arrival (AOA) \cite{GuvencWAMI15} can be employed for position estimation. In order to quantify performance bounds for such systems, several accuracy metrics are considered, including the Cram\'er-Rao lower bound (CRLB) \cite{VLP_CRLB_RSS,CRB_TOA_VLC,MFK_CRLB,RSS_aperture_JLT_2017} and the Ziv-Zakai Bound (ZZB) \cite{ZZB_MFK}.

Based on the availability of internode measurements, wireless localization networks can broadly be classified into two groups: \textit{cooperative} and \textit{noncooperative}. In the conventional noncooperative approach, position estimation is performed by utilizing only the measurements between anchor nodes (which have known locations) and agent nodes (the locations of which are to be estimated) \cite{LocatingNodes_Patwari_2005,Henk_Cooperative}. On the other hand, cooperative systems also incorporate the measurements among agent nodes into the localization process to achieve improved performance \cite{Henk_Cooperative}. Benefits of cooperation among agent nodes are more pronounced specifically for sparse networks where agents cannot obtain measurements from a sufficient number of anchors for reliable positioning \cite{BookChapter_Coop_2016}. There exists an extensive body of research regarding the investigation of cooperation techniques and the development of efficient algorithms for cooperative localization in RF-based networks (see \cite{LocatingNodes_Patwari_2005,Henk_Cooperative,BookChapter_Coop_2016} and references therein). In terms of implementation of algorithms, centralized approaches attempt to solve the localization problem via the optimization of a global cost function at a central unit to which all measurements are delivered. Among various centralized methods, maximum likelihood (ML) and nonlinear least squares (NLS) estimators are the most widely used ones, both leading to nonconvex and difficult-to-solve optimization problems, which are usually approximated through convex relaxation approaches such as semidefinite programming (SDP) \cite{SDP_Biswas_2006,SDP_Chan_2009_TSP,RSS_SDP_TVT_2010}, second-order cone programming (SOCP) \cite{SOCP_Tseng_2007,SOCP_Lampe_TWCOM_2014}, and convex underestimators \cite{Soares_Relax_TSP_2015}. In distributed algorithms, computations related to position estimation are executed locally at individual nodes, thereby reinforcing scalability and robustness to data congestion \cite{Henk_Cooperative}. Set-theoretic estimation \cite{Blatt_POCS_TSP_2006,Gholami_2011_Eurasip_CFP,Gholami_ICFP_TSP_2013,DistProj_TWCOM_2015}, factor graphs \cite{Henk_Cooperative}, and multidimensional scaling (MDS) \cite{MDS_2006} constitute common tools employed for cooperative distributed localization in the literature.

Despite the ubiquitous use of cooperation techniques in RF-based wireless localization networks, no studies in the literature have considered the use of cooperation in VLP networks. In this study, we extend the cooperative paradigm to visible light domain. More specifically, we set forth a cooperative localization framework for VLP networks whereby LED transmitters on the ceiling function as anchors with known locations and visible light communication (VLC) units with unknown locations are equipped with LEDs and photodetectors (PDs) for the purpose of communications with both LEDs on the ceiling and other VLC units. The proposed network facilitates the definition of arbitrary connectivity sets between the LEDs on the ceiling and the VLC units, and also among the VLC units, which can provide significant performance enhancements over the traditional noncooperative approach employed in the VLP literature. Based on the noncooperative (i.e., between LEDs on the ceiling and VLC units) and cooperative (i.e., among VLC units) RSS measurements, we first derive the CRLB and the MLE for the localization of VLC units. Since the MLE poses a challenging nonconvex optimization problem, we follow a set-theoretic estimation approach and formulate the problem of cooperative localization as a quasiconvex \textit{feasibility} problem (QFP) \cite{Censor_QFP_2006}, where feasible constraint sets correspond to sublevel sets of certain type of quasiconvex functions. The quasiconvexity arising in the problem formulation stems from the Lambertian formula, which characterizes the attenuation level of visible light channels. Next, we design two feasibility-seeking algorithms, having cyclic and simultaneous characteristics, which employ iterative gradient projections onto the specified constraint sets. From the viewpoint of implementation, the proposed algorithms can be implemented in a distributed architecture that relies on computations at individual VLC units and a broadcasting mechanism to update position estimates. Moreover, we provide a formal convergence proof for the projection-based algorithms based on quasi-F\'ejer convergence, which enjoys decent properties to support theoretical analysis \cite{QF_Iusem_94}.

\vspace{-0.3cm}

\subsection{Literature Survey on Set-Theoretic Estimation}
The applications of convex feasibility problems (CFPs) encompass a wide variety of disciplines, such as wireless localization \cite{Blatt_POCS_TSP_2006,SetJia_TMC_2011,Gholami_2011_Eurasip_CFP,Gholami_ICFP_TSP_2013,DistProj_TWCOM_2015}, compressed sensing \cite{Censor_CFP_SSP_2012}, image recovery \cite{Combettes_TIP_97}, image denoising \cite{ICFP_Censor_Denoising_2016} and intensity-modulated radiation therapy \cite{censor_survey_2010}. In contrary to optimization problems where the aim is to minimize the objective function while satisfying the constraints, feasibility problems seek to find a point that satisfies the constraints in the absence of an objective function \cite{Censor_CFP_SSP_2012}. Hence, the goal of a CFP is to identify a point inside the intersection of a collection of closed convex sets in a Euclidean (or, in general, Hilbert) space. In feasibility problems, a commonly pursued approach is to perform projections onto the individual constraint sets in a sequential manner, rather than projecting onto their intersection due to analytical intractability \cite{Censor_Subgradient_2008}. The work in \cite{Blatt_POCS_TSP_2006} formulates the problem of acoustic source localization as a CFP and employs the well-known projections onto convex sets (POCS) technique for convergence to true source locations. Following a similar methodology, the noncooperative wireless positioning problem with noisy range measurements is modeled as a CFP in \cite{Gholami_2011_Eurasip_CFP}, where POCS and outer-approximation (OA) methods are utilized to derive distributed algorithms that perform well under non-line-of-sight (NLOS) conditions. In \cite{SetJia_TMC_2011}, a cooperative localization approach based on projections onto nonconvex boundary sets is proposed for sensor networks, and it is shown that the proposed strategy can achieve better localization performance than the centralized SDP and the distributed MDS although it may get trapped into local minima due to nonconvexity. Similarly, the work in \cite{Gholami_ICFP_TSP_2013} designs a POCS-based distributed positioning algorithm for cooperative networks with a convergence guarantee regardless of the consistency of the formulated CFP, i.e., whether the intersection is nonempty or not.

Although CFPs have attracted a great deal of interest in the literature, QFPs have been investigated only rarely. QFPs represent generalized versions of CFPs in that the constraint sets are constructed from the lower level sets of quasiconvex functions in QFPs whereas such functions are convex in CFPs \cite{Censor_QFP_2006}. The study in \cite{Censor_QFP_2006} explores the convergence properties of subgradient projections based iterative algorithms utilized for the solution of QFPs. It is demonstrated that the iterations converge to a solution of the QFP if the quasiconvex functions satisfy H\"older conditions and the QFP is consistent, i.e., the intersection is nonempty.
In this work, we show that the Lambertian model based (originally non-quasiconvex) functions can be approximated by appropriate quasiconvex lower bounds, which convexifies the (originally nonconvex) sublevel constraint sets, thus transforming the formulated feasibility problem into a QFP.


\vspace{-0.3cm}

\subsection{Contributions}

The previous work on VLP networks has addressed the problem of position estimation based mainly on the ML estimator \cite{MFK_CRLB,RSS_aperture_JLT_2017,Guvenc_hybrid}, the least squares estimator \cite{GuvencWAMI15,Guvenc_hybrid}, triangulation \cite{zhang2014asynchronous,VLP_Accelerometer}, and trilateration \cite{TDOA_VLC} methods. This manuscript, however, considers the problem of localization in VLP networks as a feasibility problem and introduces efficient iterative algorithms with convergence guarantees in the consistent case. In addition, the theoretical bounds derived for position estimation are significantly different from those in \cite{Guvenc_hybrid,RSS_aperture_JLT_2017} via the incorporation of terms related to cooperation, which allows for the evaluation of the effects of cooperation on the localization performance in any three dimensional cooperative VLP scenario. Furthermore, unlike the previous research on localization in RF-based wireless networks via CFP modeling \cite{Blatt_POCS_TSP_2006,SetJia_TMC_2011,Gholami_2011_Eurasip_CFP,Gholami_ICFP_TSP_2013}, where a common approach is to employ POCS-based iterative algorithms, we formulate the localization problem as a QFP for VLP systems, which necessitates the development of more sophisticated algorithms (e.g., gradient projections) and different techniques for studying the convergence properties of those algorithms (e.g., quasiconvexity and quasi-Fej\'er convergence). The main contributions of this manuscript can be summarized as follows:
\begin{itemize}
	\item For the first time in the literature, we propose to employ cooperative localization for VLP networks via a generic configuration that allows for an arbitrary construction of connectivity sets and transmitter/receiver orientations.
	\item The CRLB for localization of VLC units is derived in the presence of cooperative measurements (Section~\ref{sec:systemModel}). The effects of cooperation on the performance of localization in VLP systems are illustrated based on the provided CRLB expression (Section~\ref{sec:nume_crlb}).
	\item The problem of cooperative localization in VLP systems is formulated as a quasiconvex feasibility problem, which circumvents the complexity of the nonconvex ML estimator and facilitates efficient feasibility-seeking algorithms (Section~\ref{sec:coop_cfp}).
	\item We design gradient projections based low-complexity iterative algorithms to find solutions to the feasibility problem (Section~\ref{sec:algorithm_gradient}). The proposed set-theoretic framework favors the implementation of algorithms in a distributed architecture.
	\item We provide formal convergence proofs for the proposed algorithms in the consistent case based on the concept of quasi-Fej\'er convergence (Section~\ref{sec:convergence}).
\end{itemize}

\section{System Model and Theoretical Bounds}\label{sec:systemModel}

\subsection{System Model}

\begin{figure}
\vspace{-0.2cm}
\begin{center}
\includegraphics[width=1.05\columnwidth,draft=false]{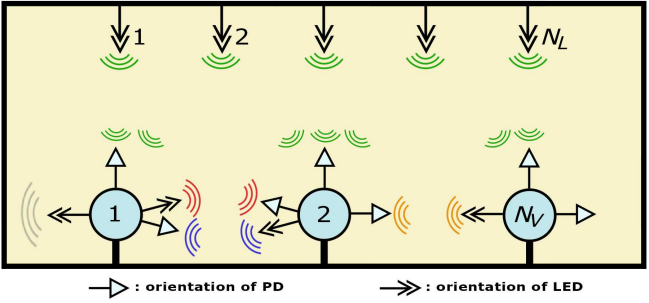}
\vspace{-0.3cm}
\caption{Cooperative VLP network.}\label{fig:system_model}
\end{center}
\vspace{-0.7cm}
\end{figure}

The proposed cooperative VLP network consists of $N_L$ LED transmitters and $N_V$ VLC units, as illustrated in Fig.~\ref{fig:system_model}. The location of the $j$th LED transmitter is denoted by $\y_{j}$ and its orientation vector is given by $\tildenTj$ for $j \in \{1,\ldots,N_L\}$. The locations and the orientations of the LED transmitters are assumed to be known, which is a reasonable assumption for practical systems \cite{Guvenc_hybrid,VLP_Accelerometer}. In the proposed system, each VLC unit not only gathers signals from the LED transmitters but also communicates with other VLC units in the system for cooperation purposes. To that aim, the VLC units are equipped with both LEDs and PDs; namely, there exist $L_i$ LEDs and $K_i$ PDs at the $i$th VLC unit for $i\in\{1,\ldots,N_V\}$. The unknown location of the $i$th VLC unit is denoted by $\ux_{i}$, where $i \in \{1, \ldots, N_V\}$. For the $j$th PD at the $i$th VLC unit, the location is given by $\ux_{i} + \ua_{i,j}$ and the orientation vector is denoted by $\uni_{R,j}$, where $j \in \{1, \ldots, K_i\}$. Similarly, for the $j$th LED at the $i$th VLC unit, the location is given by $\ux_{i} + \ub_{i,j}$ and the orientation vector is represented by $\uni_{T,j}$, where $j \in \{1, \ldots, L_i\}$. The displacement vectors, $\ua_{i,j}$'s and $\ub_{i,j}$'s, are known design parameters for the VLC units. Also, the orientation vectors for the LEDs and PDs at the VLC units are assumed to be known since they can be determined by the VLC unit design and by auxiliary sensors (e.g., gyroscope). To distinguish the LED transmitters at known locations from the LEDs at the VLC units, the former are called as the \textit{LEDs on the ceiling} (as in Fig.~\ref{fig:system_model}) in the remainder of the text.

At a given time, each PD can communicate with a subset of all the LEDs in the system. Therefore, the following connectivity sets are defined to specify the connections between the LEDs and the PDs:
\begin{align}\nonumber
\tildeSkj = \big\{ & l \in \{1,\ldots, N_L\} \mid {\textrm{$l$th LED on ceiling is}}
\\\label{eq:setStil}
&{\textrm{connected to $k$th PD of $j$th VLC unit}} \big\}
\\\nonumber
\Skij =\big\{ & l \in \{1,\ldots, L_i\} \mid {\textrm{$l$th LED of $i$th VLC unit is}}
 \\\label{eq:setS}
 &{\textrm{connected to $k$th PD of $j$th VLC unit}} \big\} .
\end{align}

\vspace{-0.2cm}

\noindent Namely, $\tildeSkj$ represents the set of LEDs on the ceiling that are connected to the $k$th PD at the $j$th VLC unit. Similarly, $\Skij$ is the set of LEDs at the $i$th VLC unit that are connected to the $k$th PD at the $j$th VLC unit.


The aim is to estimate the unknown locations, $\ux_{1},\ldots,\ux_{N_V}$, of the VLC units based on the RSS observations (measurements) at the PDs. Let $\tildePlkj$ represent the RSS observation at the $k$th PD of the $j$th VLC unit due to the transmission from the $l$th LED on the ceiling. Similarly, let $\Plkij$ denote the RSS observation at the $k$th PD of the $j$th VLC unit due to the $l$th LED at the $i$th VLC unit. Based on the Lambertian formula \cite{CRB_TOA_VLC,EPSILON}, $\tildePlkj$ and $\Plkij$ can be expressed as follows:
\vspace{-0.2cm}\begin{align}\label{eq:meas1}
\tildePlkj &= \tildealphalkj(\ux_j) + \tildenlkj \\\label{eq:meas2}
\Plkij &= \alphalkij(\ux_j,\ux_i) + \nlkij
\end{align}
where

\vspace{-0.6cm}

\begin{align}\label{eq:alp1}
\tildealphalkj(\ux_j) \triangleq -\frac{\tildeml + 1}{2\pi} \tildePTl \Akj
\frac
{\big( (\utildedlkj)^{T} \tilden_{T,l} \big)^{\tildeml}
	(\utildedlkj)^T \nRkj
}
{\norm{\utildedlkj} ^{\tildeml + 3}}
\end{align}

\vspace{-0.6cm}

\begin{align}\nonumber
&\alphalkij(\ux_j,\ux_i)  \triangleq -\frac{\mli + 1}{2\pi} \PTli \Akj\\\label{eq:alp2}
&\hspace{1.85cm}\times\frac{
	\big( (\udlkij)^{T} \nTli \big)^{\mli}
	(\udlkij)^T \nRkj }	{\norm{\udlkij}^{\mli + 3}}\,\cdot
\end{align}
for $j \in \{1, \dots, N_V\}$, $k \in \{1, \dots, K_j\}$, $i \in \{1, \dots, N_V\} \setminus  j$ and $l \in \Skij$, where $\utildedlkj \triangleq \ux_j + \ua_{j,k} - \uy_l$ and $\udlkij \triangleq \ux_j + \ua_{j,k} - \ux_i - \ub_{i,l}$. In \eqref{eq:alp1} and \eqref{eq:alp2}, $\tildeml$ ($\mli$) is the Lambertian order for the $l$th LED on the ceiling (at the $i$th VLC unit), $\Akj$ is the area of the $k$th PD at the $j$th VLC unit, $\tildePTl$ ($\PTli$) is the transmit power of the $l$th LED on the ceiling (at the $i$th VLC unit), $\tildephilkj$ ($\tildephilkij$) is the irradiation angle at the $l$th LED on the ceiling (at the $i$th VLC unit) with respect to the $k$th PD at the $j$th VLC unit, and $\tildethetalkj$ ($\tildethetalkij$) is the incidence angle for the $k$th PD at the $j$th VLC unit related to the $l$th LED on the ceiling (at the $i$th VLC unit). In addition, the noise components, $\tildenlkj$ and $\nlkij$, are modeled by zero-mean Gaussian random variables each with a variance of $\sigma^2_{j,k}$. Considering the use of a certain multiplexing scheme (e.g., time division multiplexing among the LEDs at the same VLC unit and on the ceiling, and frequency division multiplexing among the LEDs at different VLC units or on the ceiling), $\tildenlkj$ and $\nlkij$ are assumed to be independent for all different $(j,k)$ pairs and for all $l$ and $i$.


\vspace{-0.3cm}

\subsection{ML Estimator and CRLB}

Let $\ux \triangleq \big[\ux_1^T \ldots \, \ux_{N_V}^T\big]^T$ denote the vector of unknown parameters (which has a size of $3N_V \times 1$) and let $\mathbf{P}$ represent a vector consisting of all the measurements in \eqref{eq:meas1} and \eqref{eq:meas2}. The elements of $\mathbf{P}$ can be expressed as follows:
$\Bigl\{\bigl\{\bigl\{\tildePlkj
			\bigr\}_{l \in \tildeSkj}
		\bigr\}_{k \in \{1, \dots, K_j\}}
	\Bigr\}_{j \in \{1, \dots, N_V\}}, \\
	\Bigl\{\bigl\{\bigl\{\{	\Plkij\}_{l \in \Skij}
			\bigr\}_{i \in \{1, \dots, N_V\} \setminus \{j\}}
		\bigl\}_{k \in \{1, \dots, K_j\}}
	\Bigl\}_{j \in \{1, \dots, N_V\}}$.
Then, the conditional probability density function (PDF) of $\mathbf{P}$ given $\ux$, i.e., the likelihood function, can be stated as
\vspace{-0.2cm}\begin{equation}\label{eq:likelihoodSon}
	f(\mathbf{P} \,|\, \ux) =\bigg(
			\prod_{j=1}^{N_V} \prod_{k=1}^{K_j}
			\frac{1}{(\sqrt{2\pi}\, \sigma_{j,k})^{\Ntotjk}}
		\bigg)
		e^{	-\sum_{j=1}^{N_V} \sum_{k=1}^{K_j}
	 			\frac{h_{j,k}(\ux)}{2\sigma_{j,k}^2}
	 		}
\end{equation}

\vspace{-0.1cm}

\noindent where $\Ntotjk$ represents the total number of LEDs that can communicate with the $k$th PD at the $j$th VLC unit; that is, $\Ntotjk \triangleq \lvert \tildeSkj \rvert +
		\sum_{i=1,i\ne j}^{N_V} \lvert \Skij \rvert$,
and $h_{j,k}(\ux)$ is defined as
\vspace{-0.2cm}\begin{multline}\label{eq:hx}
	h_{j,k}(\ux) \triangleq
	 				\sum_{l\in \tildeSkj} \big(\tildePlkj - \tildealphalkj (\ux_j)\big)^2
	 				\
	 				\ \\
	 				+ \sum_{i=1,i\ne j}^{N_V}\,
	 				\sum_{l \in \Skij}
	 					\big( \Plkij - \alphalkij(\ux_j,\ux_i)\big)^2.
\end{multline}


\vspace{-0.1cm}

\noindent From \eqref{eq:likelihoodSon}, the maximum likelihood estimator (MLE) is obtained as
\vspace{-0.1cm}\begin{equation}\label{eq:MLE}
	\hat{\ux}_{\mathrm{ML}} = \arg\min_{\ux}
		\sum_{j=1}^{N_V} \sum_{k=1}^{K_j}
	 			\frac{h_{j,k}(\ux)}{\sigma_{j,k}^2} 
\end{equation}
and the Fisher information matrix (FIM) \cite{Poor} is given by
\begin{equation}\label{eq:FIM1}
	[\JJJ]_{t_1, t_2} = \expectation
		\bigg\{
			\frac {\partial \log f(\mathbf{P} \,|\, \ux)}
				   {\partial x_{t_1}}
			\frac {\partial \log f(\mathbf{P} \,|\, \ux)}
				   {\partial x_{t_2}}
		\bigg\}
\end{equation}
where $x_{t_1}$ ($x_{t_2}$) represents element $t_1$ ($t_2$) of vector $\ux$ with $t_1,t_2 \in \{1,2,\dots,3N_V\}$. Then, the CRLB is stated as
\vspace{-0.1cm}\begin{equation}\label{eq:CRLB}
	\rm{CRLB} = {\rm{trace}(\JJJ^{-1})} \leq \expectation\{\norm{\hat{\ux} - \ux}^2\}
\end{equation}
where $\hat{\ux}$ represents an unbiased estimator of $\ux$. From \eqref{eq:likelihoodSon} and \eqref{eq:hx}, the elements of the FIM in \eqref{eq:FIM1} can be calculated after some manipulation as
\vspace{-0.1cm}\begin{align}\nonumber
	[\JJJ]_{t_1, t_2} &= \sum_{j=1}^{N_V}\sum_{k=1}^{K_j}
\frac{1}{\sigma_{j,k}^2}
\Bigg(\sum_{l\in\widetilde{S}_k^{(j)}}
\frac{\partial\widetilde{\alpha}_{l,k}^{(j)}(\ux_j)}{\partial x_{t_1}}
\frac{\partial\widetilde{\alpha}_{l,k}^{(j)}(\ux_j)}{\partial x_{t_2}}
\\\label{eq:FIM2}
&+\sum_{i=1,i\ne j}^{N_V}\,\sum_{l\in{S}_k^{(i,j)}}
\frac{\partial{\alpha}_{l,k}^{(i,j)}(\ux_j,\ux_i)}{\partial x_{t_1}}
\frac{\partial{\alpha}_{l,k}^{(i,j)}(\ux_j,\ux_i)}{\partial x_{t_2}}\Bigg).
\end{align}

\vspace{-0.3cm}

Based on \eqref{eq:CRLB} and \eqref{eq:FIM2}, the CRLB for location estimation can be obtained for cooperative VLP systems (please see Appendix~\ref{sec:sup_partial} for the partial derivatives in \eqref{eq:FIM2}). The obtained CRLB expression is generic for any three-dimensional configuration and covers all possible cooperation scenarios via the definitions of the connectivity sets (see \eqref{eq:setStil} and \eqref{eq:setS}). To the best of authors' knowledge, such a CRLB expression has not been available in the literature for cooperative VLP systems.

\textbf{Remark~1:} \textit{From \eqref{eq:FIM2}, it is noted that the first summation term in the parentheses is related to the information from the LED transmitters on the ceiling whereas the remaining terms are due to the cooperation among the VLC units. In the noncooperative case, the elements of the FIM are given by the expression in the first line of \eqref{eq:FIM2}.}

Via \eqref{eq:CRLB} and \eqref{eq:FIM2}, the effects of cooperation on the accuracy of VLP systems can be quantified, as investigated in Section~\ref{sec:nume}.

\vspace{-0.2cm}

\section{Cooperative Localization as a Quasiconvex Feasibility Problem}\label{sec:coop_cfp}

In this section, the problem of cooperative localization in VLP networks is investigated in the framework of convex/quasiconvex feasibility. First, the feasibility approach to the localization problem is motivated, and the problem formulation is presented. Then, the convexity analysis is carried out for the resulting constraint sets.

\vspace{-0.2cm}

\subsection{Motivation}\label{sec:motivation}

For the localization of the VLC units, the MLE in \eqref{eq:MLE} has very high computational complexity as it requires a search over a $3N_V$ dimensional space. In addition, the formulation in \eqref{eq:MLE} presents a nonconvex optimization problem; hence, convex optimization tools cannot be employed to obtain the (global) optimal solution of \eqref{eq:MLE}. As the number of VLC units increases, centralized approaches obtained as solutions to a given optimization problem (such as \eqref{eq:MLE}) may become computationally prohibitive. Besides scalability issues, centralized methods also require all measurements gathered at the VLC units to be relayed to a central unit for joint processing, which may lead to communication bottlenecks. Therefore, low-complexity algorithms amenable to distributed implementation are needed to efficiently solve the cooperative localization problem in VLP networks. To that aim, the localization problem is cast as a feasibility problem with the purpose of finding a point in a finite dimensional Euclidean space that lies within the intersection of some constraint sets. Feasibility-seeking methods enjoy the advantage of not requiring an objective function, thereby eliminating the concerns for nonconvexity or nondifferentiability of the objective function \cite{Censor_SparsityFeas_2017}.
Hence, modeling the localization problem as a feasibility problem \textit{(i)} alleviates the computational burden of minimizing a (possibly nonconvex) cost function in the highly unfavorable centralized setting and \textit{(ii)} facilitates the use of efficient distributed algorithms involving parallel or sequential processing at individual VLC units.

\vspace{-0.3cm}

\subsection{Problem Formulation}\label{sec:problem_formulation}

Considering the Lambertian formula in \eqref{eq:meas1}--\eqref{eq:alp2}, an RSS measurement at a PD can be expressed as
\vspace{-0.1cm}\begin{gather}
\hat{P}_r = P_r + \eta
\end{gather}

\vspace{-0.1cm}

\noindent where $P_r$ is the true observation (as in \eqref{eq:alp1} or \eqref{eq:alp2}) and $\eta$ is the measurement noise. Suppose that the RSS measurement errors are negative, which
yields $\hat{P}_r \leq P_r$.\footnote{In order to satisfy the negative error assumption, a constant value can always be subtracted from the actual RSS measurement \cite{Tight_OA_Letter_2016}. Decreasing the value of an RSS measurement is equivalent to enlarging the corresponding feasible set. Although this assumption does not have a physical justification, it facilitates theoretical derivations and feasibility modeling of the localization problem. It will be justified via simulations in Section~\ref{sec:nume_alg} that the proposed feasibility-seeking algorithms will converge for realistic noise models (e.g., Gaussian), as well.} Then, based on \eqref{eq:alp1} and \eqref{eq:alp2}, the following inequality is obtained:
\vspace{-0.1cm}\begin{gather}\label{eq:g_x_geq}
g(\xx;\yy,\nt,\nr,m,\gamma) \leq 0
\end{gather}

\vspace{-0.1cm}

\noindent where $g: \mathbb{R}^d \rightarrow \mathbb{R}$ is the Lambertian function with respect to the unknown PD location $\xx$, defined as
\vspace{-0.1cm}\begin{gather}\label{eq:Lambertian_func}
g(\xx;\yy,\nt,\nr,m,\gamma) \triangleq \gamma - \frac{\left[(\xx-\yy)^T \nt\right]^m (\yy-\xx)^T \nr}{\norm{\xx-\yy}^{m+3}},
\end{gather}

\vspace{-0.1cm}

\noindent $\yy$, $\nt$, $\nr$, and $m$ are known, $d$ is the dimension of the visible light localization network, and $\gamma$ is given by $\gamma = \frac{\hat{P}_r}{P_t} \frac{2\pi}{(m+1)A}$.
The field-of-views (FOVs) of the LED transmitters and the PDs are taken as $90\degree$, which implies that $(\xx-\yy)^T \nt \geq 0$ and $(\yy-\xx)^T \nr \geq 0$. Under the assumption of negative measurement errors, the feasible set in which the true PD location resides is given by the following lower level set of $g(\xx)$:
\begin{gather}\label{eq:Lambertian_set}
\mtA = \Big\{ \xx \in \mathbb{R}^d ~ \Big| ~ g(\xx;\yy,\nt,\nr,m,\gamma) \leq 0  \Big\}
\end{gather}
which will hereafter be referred to as the \textit{Lambertian set}. In RF wireless localization networks, such feasible sets are generally obtained as balls \cite{Blatt_POCS_TSP_2006,Gholami_ICFP_TSP_2013}, hyperplanes \cite{Gholami_PlaneProjection_2010}, or ellipsoids \cite{Gholami_PIMRC_2010}, all of which lead to closed-form expressions for orthogonal projection. For $k \in \setfromone{K_j}$ and $j \in \setfromone{N_V}$, the Lambertian set corresponding to the $k$th PD of the $j$th VLC unit based on the signal received from the $\ell$th LED on the ceiling for $\ell \in \tildeSkj$ is defined as follows:
\begin{gather}\label{eq:Lambertian_set_noncoop}
\mtN_{\ell,k}^{(j)} = \Big\{ \zz \in \mathbb{R}^d ~ \Big| ~ \tilde{g}_{\ell,k}^{(j)}(\zz) \leq 0 \Big\}
\end{gather}
where $\tilde{g}_{\ell,k}^{(j)}(\zz)$ is given by
\begin{gather}\label{eq:Lambert_func_noncoop}
\tilde{g}_{\ell,k}^{(j)}(\zz) \triangleq g\big(\zz;\yy_{\ell}-\ua_{j,k},\tildenTjgen{\ell},\nRkjgen{k}{j},\tildemlgen{\ell},\gammatildelkj\big)
\end{gather}
and $\gammatildelkj$ is calculated from \eqref{eq:meas1}. Similarly, the Lambertian set corresponding to the $k$th PD of the $j$th VLC unit based on the signal received from the $\ell$th LED of the $i$th VLC unit for $\ell \in \Skij$ is defined as
\begin{gather}\label{eq:Lambertian_set_coop}
\mtC_{\ell,k}^{(i,j)} = \Big\{ \zz \in \mathbb{R}^d ~ \Big| ~ g_{\ell,k}^{(i,j)}(\zz,\xx_i)  \leq 0 \Big\}
\end{gather}
where $g_{\ell,k}^{(i,j)}(\zz,\xx_i)$ is given by
\begin{gather}\label{eq:Lambert_func_coop}
g_{\ell,k}^{(i,j)}(\zz,\xx_i) \triangleq g\big(\zz;\ux_i + \ub_{i,\ell}-\ua_{j,k},\unTgen{i}{\ell},\nRkjgen{k}{j},\mligen{\ell}{i},\gammalkij\big)
\end{gather}
and $\gammalkij$ is calculated from \eqref{eq:meas2}. The sets defined as in \eqref{eq:Lambertian_set_noncoop} represent noncooperative localization as they are constructed from the RSS measurements corresponding to the LEDs on the ceiling, whereas the sets in \eqref{eq:Lambertian_set_coop} are based on the signals from the LEDs of the other VLC units and represent the cooperation among the VLC units. Assuming negatively biased RSS measurements, the problem of cooperative localization in a visible light network reduces to that of finding a point in the intersection of sets as defined in \eqref{eq:Lambertian_set_noncoop} and \eqref{eq:Lambertian_set_coop} for each VLC unit. If the Lambertian function in \eqref{eq:Lambertian_func} is assumed to be quasiconvex\footnote{The conditions under which the Lambertian function is quasiconvex are investigated in Section~\ref{sec:convexity_Lambertian}.}, then the \textit{quasiconvex feasibility problem} (QFP) can be formulated as follows \cite{Bauschke_Proj_ConvFeas_1996,Censor_QFP_2006}:

	\textbf{Problem~1:} \textit{Let $\xx \triangleq \left( \xx_1, \ldots, \xx_{N_V} \right)$. The feasibility problem for cooperative localization of VLC units is given by\footnote{
			It may be more convenient to regard the problem in \eqref{eq:icfp} as an \textit{implicit} quasiconvex feasibility problem (IQFP) since the Lambertian sets $\mtC_{\ell,k}^{(i,j)}$ depend on the locations of the VLC units, which are not known \textit{a priori} \cite{Gholami_ICFP_TSP_2013}.
			It should be emphasized that the feasibility problem posed in Problem~1 is different from those in RF-based localization systems (e.g., \cite{Gholami_2011_Eurasip_CFP,Gholami_ICFP_TSP_2013}) since the constraint sets and the associated quasiconvex functions have distinct characteristics as compared to convex functions (e.g., distance to a ball) encountered in RF-based systems.}}
		\begin{align}\nonumber
			\mathop{\mathrm{find}}~~ &\xx \in \mathbb{R}^{d N_V}
			\\ \label{eq:icfp}
			\mathrm{subject~to}~~ &\xx_j \in \Lambda_j \cap \Upsilon_j ,\, j = 1,\ldots,N_V
		\end{align}
		\textit{where}
		\vspace{-0.4cm}\begin{align}\label{eq:set_all_noncoop}
		\Lambda_j &= \bigcap_{k=1}^{K_j} \bigcap_{\ell \in \tildeSkj} \mtN_{\ell,k}^{(j)}
		\\\label{eq:set_all_coop}
		\Upsilon_j &= \bigcap_{k=1}^{K_j} \bigcap_{i=1}^{N_V} \bigcap_{\ell \in \Skij} \mtC_{\ell,k}^{(i,j)}~.
		\end{align}

\vspace{-0.2cm}

\subsection{Convexity Analysis of Lambertian Sets}\label{sec:convexity_Lambertian}

The Lambertian sets as defined in \eqref{eq:Lambertian_set} are not convex in general. The following lemma presents the conditions under which the Lambertian sets become convex.

	\textbf{Lemma~1:} \textit{Consider the $\alpha$-level set}
	\begin{align}\label{eq:lemma1_set}
	\mtB = \Big\{ \xx \in \Omega ~ \Big| ~ g_{\epsilon}(\xx) \leq \alpha \Big\}
	\end{align}
	\textit{of $g_{\epsilon}(\xx)$, which is given by}
	\begin{align}\label{eq:lemma1_func}
	g_{\epsilon}(\xx) = \gamma - \frac{(\yy-\xx)^T \nr}{\norm{\xx-\yy}^k + \epsilon}
	\end{align}
	\textit{where $\epsilon$ is a small positive constant to avoid non-differentiability and non-continuity of $g_{\epsilon}(.)$ at $\yy$, as in \cite[Eq. 7]{Mathlouthi2016}, $k \geq 1$ and $\gamma > 0$ are real numbers, and $\Omega \subset \mathbb{R}^d$ is defined as}
	\begin{align}\label{eq:lemma1_subset}
	\Omega = \Big\{ \xx \in \mathbb{R}^d ~ \Big| ~ (\yy-\xx)^T \nr \geq 0  \Big\}~.
	\end{align}
	\textit{Then, $\mtB$ is convex for each $\alpha \in \mathbb{R}$.}

\indent\textit{Proof}: Please see Appendix~\ref{sec:sup_lemma1}.

	\textbf{Remark~2:} \textit{Lemma~1 characterizes the type of Lambertian functions whose sublevel sets are convex. Since a function whose all sublevel sets are convex is quasiconvex \cite{NP_Book_2006}, Lambertian functions of the form \eqref{eq:lemma1_func} are quasiconvex over the halfspace $\Omega$ in \eqref{eq:lemma1_subset}. It can be noted that $\Omega$ consists of those VLC unit locations which are able to obtain measurements from an LED located at $\yy$ due to the receiver FOV limit of $90 \degree$.}


\vspace{-0.2cm}

\subsection{Convexification of Lambertian Sets}

In this part, we utilize Lemma~1 to investigate the following two cases in which the Lambertian functions can be transformed into the form of \eqref{eq:lemma1_func} and Problem~1 becomes a QFP.

\subsubsection{Case 1: Convexification via Majorization}\label{sec:expanded_Lambertian}

We propose to approximate the Lambertian function $g(\xx)$ in \eqref{eq:Lambertian_func} by a \textit{quasiconvex minorant} $\tilde{g}(\xx)$ such that $\tilde{g}(\xx) \leq g(\xx)$ for $\xx \in \Omega$ and $\mtA \subseteq \tilde{\mathcal{L}}$, where $\tilde{\mathcal{L}} \triangleq \big\{ \xx \in \Omega ~ \big| ~ \tilde{g}(\xx) \leq 0  \big\}$ represents a majorization of the original set $\mtA \triangleq \big\{ \xx \in \Omega ~ \big| ~ g(\xx) \leq 0  \big\}$. Assuming $\xx \in \Omega$, we have
\begin{align}\label{eq:case1_start}
g(\xx) &= \gamma - \frac{\left[(\xx-\yy)^T \nt\right]^m (\yy-\xx)^T \nr}{\norm{\xx-\yy}^{m+3}} \\ \label{eq:case1_1}
&\geq \gamma - \frac{\norm{\xx-\yy}^m \norm{\nt}^m (\yy-\xx)^T \nr}{\norm{\xx-\yy}^{m+3}}
\\ \label{eq:case1_2}
&= \gamma - \frac{(\yy-\xx)^T \nr}{\norm{\xx-\yy}^3} \triangleq \tilde{g}(\xx)
\end{align}
where \eqref{eq:case1_1} is due to the Cauchy-Schwarz inequality and $\xx \in \Omega$, and \eqref{eq:case1_2} follows from the unit norm property of the orientation vector. Then, including $\epsilon$ in the denominator, we construct the Lambertian sets as (hereafter called \textit{expanded Lambertian sets})
\begin{gather}\label{eq:Lambertian_set_expanded}
\mtA = \Big\{ \xx \in \Omega ~ \Big| ~ \tilde{g}_{\epsilon}(\xx) \leq 0  \Big\}
\end{gather}
with
\vspace{-0.2cm}\begin{gather}\label{eq:quasiconvex_minorant}
\tilde{g}_{\epsilon}(\xx) = \gamma - \frac{(\yy-\xx)^T \nr}{\norm{\xx-\yy}^3 + \epsilon}
\end{gather}
and $\Omega$ being as in \eqref{eq:lemma1_subset}. According to Lemma~1, $\mtA$ in \eqref{eq:Lambertian_set_expanded} is convex, $\tilde{g}_{\epsilon}(\xx)$ in \eqref{eq:quasiconvex_minorant} is quasiconvex over $\Omega$ and the resulting problem of determining a point inside the intersection of such sets turns into a QFP, which can be studied through iterative projection algorithms \cite{Bauschke_Proj_ConvFeas_1996,Censor_QFP_2006}. 

\subsubsection{Case 2: Known VLC Height, Perpendicular LED}\label{sec:known_height_perp_LED}

In this case, as in \cite{CRB_TOA_VLC,VLP_CRLB_RSS,LED_MultiRec,MFK_CRLB}, it is assumed that the LED transmitters on the ceiling have perpendicular orientations, i.e., $\tildenTj = \left[ 0 ~~  0  ~ -1 \right]^T$ for each $j \in \setfromone{N_L}$, and the height of each VLC unit is known. This assumption is valid for some practical scenarios, an example of which is a VLP network where the LEDs on the ceiling are pointing downwards and the VLC units are attached to robots that move over a two-dimensional plane \cite[Fig.~3]{VLP_Roadmap}. Assuming that the height of the LED transmitters relative to the VLC units is $h$ and $\nt = \left[ 0 ~~  0  ~ -1 \right]^T$, the Lambertian function in \eqref{eq:Lambertian_func} can be rewritten as follows:
\begin{gather}\label{eq:Lambertian_func_case2}
g(\xx;\yy,\nt,\nr,m,\gamma) = \gamma - \frac{h^m (\yy-\xx)^T \nr}{\norm{\xx-\yy}^{m+3}}\,\cdot
\end{gather}
Then, the Lambertian set corresponding to the function in \eqref{eq:Lambertian_func_case2} by introducing $\epsilon$ in the denominator is obtained as
\begin{gather}\label{eq:Lambertian_set_case2}
\mtA = \Big\{ \xx \in \Omega ~ \Big| ~ \tilde{g}_{\epsilon}(\xx)  \leq 0  \Big\}
\end{gather}
with
\vspace{-0.2cm}\begin{gather}
\tilde{g}_{\epsilon}(\xx) = \tilde{\gamma} - \frac{(\yy-\xx)^T \nr}{\norm{\xx-\yy}^{m+3} + \epsilon}
\end{gather}
where $\Omega$ is given by \eqref{eq:lemma1_subset} and $\tilde{\gamma} = \gamma/h^m$. Note that the Lambertian set in \eqref{eq:Lambertian_set_case2} is effectively defined on $\mathbb{R}^2$ since the height of the VLC unit is already known. According to Lemma~1, the set defined in \eqref{eq:Lambertian_set_case2} is convex. Therefore, in this case, the noncooperative sets as defined in \eqref{eq:Lambertian_set_noncoop} are originally convex.

Based on the discussion above, it is concluded that in the case of a known VLC height and perpendicular LED transmitter orientations, the expanded Lambertian sets in \eqref{eq:Lambertian_set_expanded} defined on $\mathbb{R}^2$ must be used for the measurements among the VLC units (i.e., cooperative measurements) in order to ensure that Problem~1 is a QFP. For the general case in which the LED orientations are arbitrary and/or the heights of the VLC units are unknown, all the noncooperative and cooperative Lambertian sets must be replaced by the corresponding expanded versions in \eqref{eq:Lambertian_set_expanded}.

\begin{figure}
    \vspace{-0.5cm}
	\begin{center}		
		\subfigure[]{
			\label{fig:VLP_Network_Noncoop}
			\includegraphics[scale=0.75]{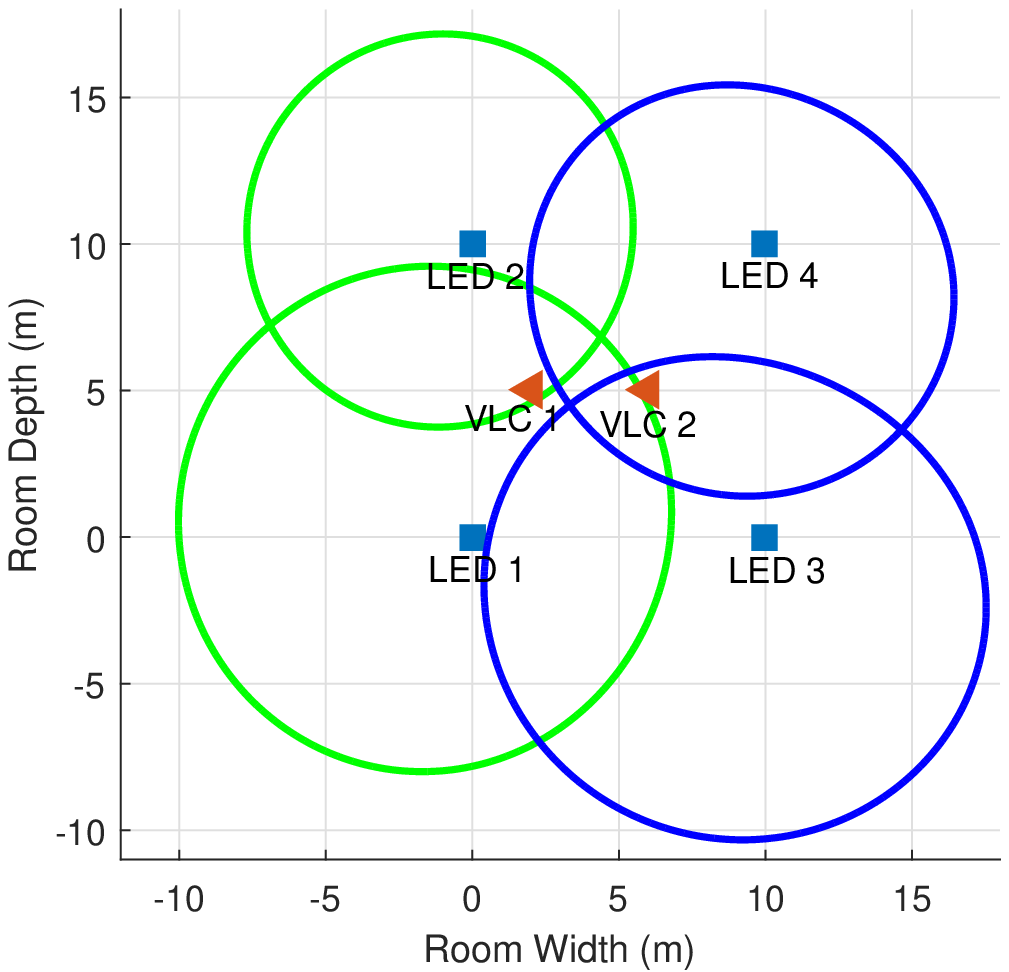}
		}
		\vspace{-0.4cm}
		\subfigure[]{
			\label{fig:Coop_Sets_Labeled_Zoomed}
			\includegraphics[scale=0.75]{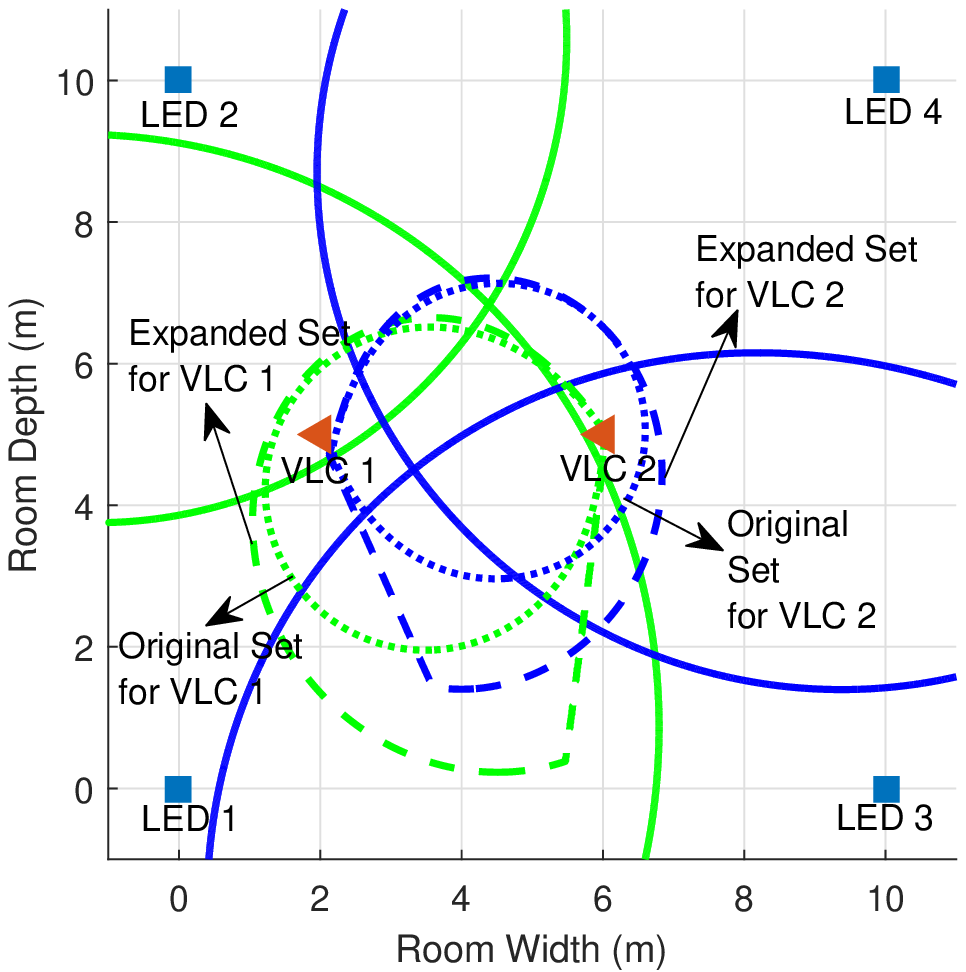}
		}
	\end{center}
	\caption{\subref{fig:VLP_Network_Noncoop} A noncooperative VLP network consisting of four LED transmitters on ceiling and two VLC units. VLC-$1$ is connected to LED-$1$ and LED-$2$, and VLC-$2$ is connected to LED-$3$ and LED-$4$. Green and blue regions represent the noncooperative Lambertian sets for VLC-$1$ and VLC-$2$, respectively. \subref{fig:Coop_Sets_Labeled_Zoomed} Cooperative version of the VLP system in Fig.~\ref{fig:VLP_Network_Noncoop}, shown by zooming onto VLC units. Case 1 type expanded cooperative Lambertian sets and their nonexpanded (original) counterparts are illustrated along with noncooperative Lambertian sets. Cooperation helps shrink the intersection region of Lambertian sets for VLC units.}
	\label{fig:CoopVLP_ExplanatoryFigure}
\end{figure}

A noncooperative VLP network is illustrated in Fig.~\ref{fig:VLP_Network_Noncoop}, where there exist four LED transmitters on the ceiling and two VLC units. In the network, it is assumed that the heights of the VLC units are known and the LEDs on the ceiling have perpendicular orientations so that Case~2 type convex Lambertian sets can be utilized for the measurements between the LEDs on the ceiling and the VLC units. Fig.~\ref{fig:Coop_Sets_Labeled_Zoomed} shows the cooperative version of the VLP network with cooperative Lambertian sets including both the nonexpanded (original) sets as in \eqref{eq:Lambertian_set_coop} and Case~1 type expanded sets as in \eqref{eq:Lambertian_set_expanded}. It is noted from Fig.~\ref{fig:Coop_Sets_Labeled_Zoomed} that incorporating cooperative Lambertian sets into the localization geometry can significantly reduce the region of intersection of the Lambertian sets.

\section{Gradient Projections Algorithms}\label{sec:algorithm_gradient}

In this section, we design iterative subgradient projections based algorithms to solve Problem~1. The idea of using subgradient projections is to approach a convex set defined as a lower contour set of a convex/quasiconvex function by moving in the direction that decreases the value of that function at each iteration, i.e., in the opposite direction of the subgradient of the function at the current iterate \cite{Censor1982_CSP,Censor_Subgradient_2008}. First, the definition of the gradient projector is presented as follows:

	\textbf{Definition~1:} The gradient projection operator $G_f^{\lambda} : \mathbb{R}^d \rightarrow \mathbb{R}^d$ onto the zero-level set of a continuously differentiable function $f: \mathbb{R}^d \rightarrow \mathbb{R}$ is given by \cite{QuasiFejer_Combettes_2001}
	\vspace{-0.1cm}\begin{equation}\label{eq:gradient_projector}
	\Gflambda{f}{\lambda} : \xx \mapsto \begin{cases}
	\xx - \lambda \frac{f(\xx)}{\norm{\nabla f(\xx)}^2} \nabla f(\xx), & \text{if $f(\xx) > 0$} \\
	\xx, & \text{if $f(\xx) \leq 0$}
	\end{cases}
	\end{equation}
	where $\lambda$ is the relaxation parameter and $\nabla$ is the gradient operator. The gradient projector can also be expressed as
	\vspace{-0.1cm}\begin{align}\label{eq:gradient_projector2}
	\Gflambda{f}{\lambda}(\xx) = \xx - \lambda \frac{f^{+}(\xx)}{\norm{\nabla f(\xx)}^2} \nabla f(\xx)
	\end{align}
	with $f^{+}(\xx)$ denoting the positive part, i.e., $f^{+}(\xx) = \max\{ 0, f(\xx) \}$. In the sequel, it is assumed that $\Gflambda{f}{\lambda}(\xx) = \xx$ when $\xx$ is outside the region where $f$ is quasiconvex.

\vspace{-0.3cm}

\subsection{Projection Onto Intersection of Halfspaces}\label{sec:proj_halfspace}

Since the functions of the form \eqref{eq:lemma1_func} are continuously differentiable and quasiconvex on the halfspace $\Omega$ in \eqref{eq:lemma1_subset}, a special case of subgradient projections, namely, gradient projections, can be utilized to solve Problem~1, under the constraint that iterates must be inside $\Omega$ to guarantee quasiconvexity. Hence, at the start of each iteration of gradient projections, projections onto the intersection of halfspaces of the form $\Omega$ in \eqref{eq:lemma1_subset} can be performed to keep the iterates inside the quasiconvex region. The procedure for projection onto the intersection of halfspaces
\vspace{-0.3cm}\begin{align}\label{eq:polyhedron}
\Gamma_j = \bigcap_{k=1}^{K_j} \bigcap_{\ell \in \tildeSkj} \tilde{\Omega}_{\ell,k}^{(j)}
\end{align}
corresponding to the $j$th VLC unit for $j \in \setfromone{N_V}$, with the halfspaces given by
\vspace{-0.1cm}\begin{align}\label{eq:halfspace}
\tilde{\Omega}_{\ell,k}^{(j)} \triangleq \Big\{ \xx \in \mathbb{R}^d ~ \Big| ~ (\yy_{\ell}-\ua_{j,k}-\xx)^T \nRkjgen{k}{j} \geq 0  \Big\} ~,
\end{align}
is provided in Algorithm~\ref{alg:polyhedron}\footnote{$P_{\mathcal{C}}(\xx)$ denotes the orthogonal projection operator, i.e., $P_{\mathcal{C}}(\xx) = \argmin_{\ww \in \mathcal{C}} \norm{\ww - \xx}$.}. In order to find a point inside the intersection of halfspaces, the method of alternating (cyclic) projections is employed in Algorithm~1, where the current iterate is projected onto each halfspace in a cyclic manner. Convergence properties of this method are well studied in the literature \cite{vonNeumann_MAP,Halperin_1962_Product}. $\Gamma_j$ is guaranteed to be nonempty since it represents the set of possible locations for the $j$th VLC unit at which the RSS measurements from the connected LEDs on the ceiling can be acquired. However, the intersection of the halfspaces corresponding to the LEDs of the other VLC units that are connected to the $j$th VLC unit may be empty due to the VLC unit locations being unknown and variable during iterations.

\vspace{-0.1cm}

\begin{algorithm}
	\caption{Projection Onto Intersection of Halfspaces $\Gamma_j$}
	\label{alg:polyhedron}
	\footnotesize
	\begin{algorithmic}
		\Function{$P_{\Gamma_j}$}{$\xx_j$}
		\State \textbf{Initialization:} $\xx_j^{(0)} = \xx_j$
		\State \textbf{Iterative Step:} Given the $n$th iterate $\xx_j^{(n)} \in \mathbb{R}^d$
		\For{$k = 1,\ldots,K_j$}
		\For{$\ell \in \tildeSkj$}
		\vspace{-0.3cm}
		\begin{equation}
		\xx_j^{(n)} = P_{\tilde{\Omega}_{\ell,k}^{(j)}} (\xx_j^{(n)})
		\vspace{-0.3cm}
		\end{equation}
		\EndFor
		\EndFor
		\State
		Set $\xx_j^{(n+1)} = \xx_j^{(n)}$
		\State \textbf{Stopping Criterion:} $\norm{\xxit{j}{n+1} - \xxit{j}{n}} < \delta$ for some $\delta > 0$.
		\EndFunction
	\end{algorithmic}
	\normalsize
\end{algorithm}

\vspace{-0.65cm}

\subsection{Step Size Selection}\label{sec:step_size}

An important phase of the proposed projection algorithms is determining the relaxation parameters (i.e., step sizes) associated with the gradient projector. The step size selection procedure exploits the well-known Armijo rule, which is an inexact line search method used extensively for gradient descent methods in the literature \cite{Armijo_1966,Wolfe_1969},\cite[Section 1.2]{nonlin_program_Bertsekas_99}. Algorithm~\ref{alg:armijo} provides an Armijo-like procedure for step size selection given a set of Lambertian functions, the initial step size value $\lambda$, a fixed constant $\beta \in (0,1)$ specifying the degree of decline in the value of the function, step size shrinkage factor $\xi \in (0,1)$, and the current point $\xx$. The guarantee of existence of a step size as described in Algorithm~\ref{alg:armijo} can be shown similarly to \cite[Lemma 4]{Fliege2000}.

\vspace{-0.1cm}

\begin{algorithm}
	\caption{Armijo Rule for Step Size Selection}
	\label{alg:armijo}
\footnotesize
	\begin{algorithmic}
		\Function{$\mathcal{J}$}{$\{f_i\}_{i=1}^M, \lambda, \beta, \xi, \xx$}
		\State \textbf{Output:} New step size $\tilde{\lambda}$
		\State Set the step size as
		\vspace{-0.1cm}
		\begin{equation} \label{eq:armijo_step}
		\tilde{\lambda} = \lambda \xi^{\tilde{m}}
		\vspace{-0.3cm}
		\end{equation}
		where	
				\vspace{-0.2cm}
		\begin{align}\nonumber
		\tilde{m} &= \min\{ m \in \mathbb{Z}_{\geq 0} ~ | \\ \label{eq:armijo_selection}
		&f_i( \Gflambda{f_i}{\lambda \xi^m}(\xx) ) \leq f_i(\xx) (1 - \beta \lambda \xi^m),\, \forall i \in \setfromone{M} \}
		\end{align}
		\vspace{-0.3cm}
		\EndFunction
	\end{algorithmic}
	\normalsize
\end{algorithm}

\vspace{-0.65cm}

\subsection{Iterative Projection Based Algorithms}\label{sec:algorithms}

In this work, two classes of gradient projections algorithms, namely, sequential (i.e., cyclic) \cite{Censor1982_CSP} and simultaneous (i.e., parallel) \cite{DosSantos1987} projections, are considered for the QFP described in Problem~1. The proposed algorithm for cyclic projections, namely, the cooperative cyclic gradient projections (CCGP) algorithm, for cooperative localization of VLC units is provided in Algorithm~\ref{alg:cyclic}. In the proposed cyclic projections, the current iterate, which signifies the location of the given VLC unit, is first projected onto the intersection of halfspaces corresponding to the LEDs on the ceiling via Algorithm~\ref{alg:polyhedron}. Then, the resulting point is projected onto the noncooperative Lambertian set that leads to the highest function value, i.e., the most violated constraint set \cite{Censor_QFP_2006}. Similarly, projection onto the most violated constraint set among the cooperative Lambertian sets is performed and the projections obtained by noncooperative and cooperative sets are weighted to obtain the next iterate.

The cooperative simultaneous gradient projections (CSGP) algorithm is proposed as detailed in Algorithm~\ref{alg:sim}. Simultaneous projections are based on projecting the current point onto each noncooperative and cooperative Lambertian set separately and then averaging all the resulting points to obtain the next iterate. At each iteration, the parallel projection stage is preceded by projection onto the intersection of halfspaces, which aims to ensure that the current iterate resides in the region where all the Lambertian functions corresponding to the fixed anchors (i.e., the LEDs on the ceiling) are quasiconvex. It should be noted that for both cyclic and simultaneous projections, the cooperative Lambertian sets are determined by the latest estimates of the VLC unit locations \cite{Gholami_ICFP_TSP_2013}, which are updated in the ascending order of their indices. In addition, the step sizes are updated using the Armijo rule in Algorithm~\ref{alg:armijo}.

\textbf{Remark~3:} \textit{Both Algorithm~\ref{alg:cyclic} and Algorithm~\ref{alg:sim} can be implemented in a distributed manner by employing a gossip-like procedure among the VLC units \cite{boyd2006randomized}. After refining its location estimate via projection methods, each VLC unit broadcasts the resulting updated location to other VLC units to which it is connected. In order to save computation time, a synchronous counterpart of this asynchronous/sequential algorithm can be devised, where VLC units work in parallel to update their locations based on the most recent broadcast information. Hence, the synchronous/parallel implementation trades off the localization accuracy for faster convergence to the desired solution.}

\begin{algorithm}
	\caption{Cooperative Cyclic Gradient Projections (CCGP)}
	\label{alg:cyclic}
	\footnotesize
	\begin{algorithmic}
		\State \textbf{Initialization:} Choose an arbitrary initial point $\left(\xxit{1}{0},\ldots,\xxit{N_V}{0}\right) \in \mathbb{R}^{d N_V}$.
		\State \textbf{Iterative Step:} Given the $n$th iterate $\left(\xxit{1}{n},\ldots,\xxit{N_V}{n}\right) \in \mathbb{R}^{d N_V}$
		\For{$j = 1,\ldots,N_V$}
		\State \textit{Projection Onto Intersection of Halfspaces $\Gamma_j$ by Algorithm~\ref{alg:polyhedron}:}
		\vspace{-0.2cm}
		\begin{align}\label{eq:proj_polyhedron_cyclic}
		\tilde{\xx}_{j}^{(n)} = P_{\Gamma_j}(\xxit{j}{n})
		\end{align}
		\vspace{-0.3cm}
		\State \textit{Most Violated Constraint Control for Noncooperative Projections:}
		\vspace{-0.1cm}
		\begin{align}\label{eq:mvcc_noncoop}
		(\hat{k}_{\rm{nc}},\hat{\ell}_{\rm{nc}}) = \arg \max_{k,\ell} \glkjfunc\big(\tildexjn\big)
		\end{align}
		\vspace{-0.1cm}
		
		\State \textit{Most Violated Constraint Control for Cooperative Projections:}
		\vspace{-0.1cm}
		\begin{align}\label{eq:mvcc_coop}
		(\hat{k}_{\rm{c}},\hat{i}_{\rm{c}},\hat{\ell}_{\rm{c}}) = \arg \max_{k,i,\ell} \mtG_j^{(n)}
		\end{align}
		where
        \vspace{-0.53cm}
		\begin{align}
		\mtG_j^{(n)} \triangleq \Big \{ \glkijfunc(\tildexjn,\ux_i^{(\hat{n})}) ~ \Big| ~ \tildexjn \in \Omega_{\ell, k}^{(i,j)} \Big \}
		\end{align}
		\vspace{-0.4cm}
		\begin{align}\label{eq:halfspace_coop}
		\Omega_{\ell, k}^{(i,j)} \triangleq \Big\{ \xx \in \mathbb{R}^d ~ \Big| ~ (\ux_i^{(\hat{n})} + \ub_{i,\ell}-\ua_{j,k}-\xx)^T \nRkjgen{k}{j} \geq 0  \Big\}
		\vspace{-0.1cm}
		\end{align}	
		with $\hat{n} = n$ for $i > j$, $\hat{n} = n+1$ for $i < j$.
		
		\State \textit{Averaging:}
		\vspace{-0.1cm}
		\begin{align}\label{eq:cyclic_iterations}
		\xxit{j}{n+1} = \vartheta_{\rm{nc}} \Gflambda{\tilde{g}_{\hat{\ell}_{\rm{nc}},\hat{k}_{\rm{nc}}}^{(j)}}{\lambda_{j,\rm{nc}}^{(n)}}(\tildexjn) + \vartheta_{\rm{c}} \Gflambda{g_{\hat{\ell}_{\rm{c}},\hat{k}_{\rm{c}}}^{(\hat{i}_{\rm{c}},j)}(.,\ux_i^{(\hat{n})})}{\lambda_{j,\rm{c}}^{(n)}}(\tildexjn)
		\vspace{-0.1cm}
		\end{align}
		where $\vartheta_{\rm{nc}} + \vartheta_{\rm{c}} = 1$ and $\vartheta_{\rm{nc}} \geq 0$, $\vartheta_{\rm{c}} \geq 0$.
		\EndFor
		
		\State \textbf{Stopping Criterion:} $\sum_{j=1}^{N_V} \norm{\xxit{j}{n+1} - \xxit{j}{n}}^2 < \delta$ for some $\delta > 0$.
		
		\State \textbf{Relaxation Parameters:} Initialize $\lambda_{j,\rm{nc}}^{(0)} = \lambda_{j,\rm{c}}^{(0)} = \lambda_0$ and update using Algorithm~\ref{alg:armijo} as
		\vspace{-0.1cm}
		\begin{align}\label{eq:step_size_cyc_nc}
		\lambda_{j,\rm{nc}}^{(n)} = \mathcal{J}(\tilde{g}_{\hat{\ell}_{\rm{nc}},\hat{k}_{\rm{nc}}}^{(j)}, \lambda_{j,\rm{nc}}^{(n-1)}, \beta, \xi, \tildexjn)
		\end{align}	
		\vspace{-0.1cm}	
		\begin{equation}\label{eq:step_size_cyc_c}
		\lambda_{j,\rm{c}}^{(n)} = \begin{cases}
		\mathcal{J}(g_{\hat{\ell}_{\rm{c}},\hat{k}_{\rm{c}}}^{(\hat{i}_{\rm{c}},j)}(.,\ux_i^{(\hat{n})}), \lambda_{j,\rm{c}}^{(n-1)}, \beta, \xi, \tildexjn), & \text{if $\mtG_j^{(n)} \neq \emptyset$} \\
		\lambda_{j,\rm{c}}^{(n-1)} & \text{otherwise}
		\end{cases}
		\vspace{-0.1cm}
		\end{equation}
		for $j \in \setfromone{N_V}$.
	\end{algorithmic}
	\normalsize
\end{algorithm}



\begin{algorithm}
	\caption{Cooperative Simultaneous Gradient Projections (CSGP)}
	\label{alg:sim}
	\footnotesize
	\begin{algorithmic}
		\State \textbf{Initialization:} Choose an arbitrary initial point $\left(\xxit{1}{0},\ldots,\xxit{N_V}{0}\right) \in \mathbb{R}^{d N_V}$.
		\State \textbf{Iterative Step:} Given the $n$th iterate $\left(\xxit{1}{n},\ldots,\xxit{N_V}{n}\right) \in \mathbb{R}^{d N_V}$
		\For{$j = 1,\ldots,N_V$}
		\State \textit{Projection Onto Intersection of Halfspaces $\Gamma_j$ by Algorithm~\ref{alg:polyhedron}:}
		\vspace{-0.1cm}
		\begin{align}\label{eq:proj_polyhedron}
		\tilde{\xx}_{j}^{(n)} = P_{\Gamma_j}\big(\xxit{j}{n}\big)
		\end{align}
		\State \textit{Parallel Projection Onto Lambertian Sets:}
		\vspace{-0.1cm}
		\begin{align} \nonumber
		\xxit{j}{n+1} = \sum_{k=1}^{K_j} & \Bigg[ \sum_{\ell \in \tildeSkj } \tildekappa \Gflambda{\tilde{g}_{\ell,k}^{(j)}}{\lambda_j^{(n)}}(\tilde{\xx}_{j}^{(n)}) \\ \label{eq:sim_proj} &+ \sum_{i=1,i \neq j}^{N_V} \sum_{\ell \in \Skij} \kappalkij \Gflambda{g_{\ell,k}^{(i,j)}(.,\xxit{i}{\hat{n}})}{\lambda_j^{(n)}}(\tilde{\xx}_{j}^{(n)})   \Bigg]
		\end{align}
		where $\hat{n} = n$ for $i > j$, $\hat{n} = n+1$ for $i < j$ and the weights satisfy
		\vspace{-0.1cm}
		\begin{align} \label{eq:weights}
		\sum_{k=1}^{K_j} \left( \sum_{\ell \in \tildeSkj } \tildekappa +  \sum_{i=1,i \neq j}^{N_V} \sum_{\ell \in \Skij} \kappalkij \right) = 1
		\end{align}
		and $\tildekappa \geq 0$, $\kappalkij \geq 0$, $\forall i,\ell,k$.
		\EndFor
		\State \textbf{Stopping Criterion:} $\sum_{j=1}^{N_V} \norm{\xxit{j}{n+1} - \xxit{j}{n}}^2 < \delta$ for some $\delta > 0$.
		\State \textbf{Relaxation Parameters:} Initialize $\lambda_j^{(0)} = \lambda_0$ and update using Algorithm~\ref{alg:armijo} as
		\vspace{-0.1cm}
		\begin{equation}\label{eq:alg_sim_step_size}
		\lambda_j^{(n)} = \mathcal{J}(\mtFtilde_j \cup \mtS_j^{(n)}, \lambda_j^{(n-1)}, \beta, \xi, \tildexjn)
		\vspace{-0.1cm}
		\end{equation}
		for $j \in \setfromone{N_V}$, where $\mtFtilde_j$ and $\mtF_j$ are given by \eqref{eq:Ftildej} and \eqref{eq:Fj} in Appendix~\ref{sec:sup_prop1}, respectively, and
		\vspace{-0.1cm}
		\begin{equation}
		\mtS_j^{(n)} \triangleq \big\{ f \in \mtF_j ~ |~ f(\tildexjn) \leq \gammalkij \big\}.
		\vspace{-0.1cm}
		\end{equation}
	\end{algorithmic}
	\normalsize
\end{algorithm}



\vspace{-0.2cm}

\section{Convergence Analysis}\label{sec:convergence}

In this section, the convergence analysis of the proposed algorithms in Algorithm~\ref{alg:cyclic} and Algorithm~\ref{alg:sim} is performed in the consistent case. To that aim, it is assumed that for each $j \in \setfromone{N_V}$, the intersection of the noncooperative and cooperative Lambertian sets in \eqref{eq:icfp} is nonempty; that is, $\Lambda_j \cap \Upsilon_j \neq \emptyset$, where $\Lambda_j$ and $\Upsilon_j$ are given by \eqref{eq:set_all_noncoop} and \eqref{eq:set_all_coop}, respectively. In the following, we present the definitions of quasiconvexity and quasi-Fej\'er convergence, which will be used for the convergence proofs.

	\textbf{Definition~2} \textit{(Quasiconvexity \cite{mangasarian1994nonlinear}):} A differentiable function $f: \mathbb{R}^n \rightarrow \mathbb{R}$ is quasiconvex if and only if $f(\xx) \leq f(\yy)$ implies $\nabla f(\yy)^T (\xx - \yy) \leq 0 ~ \forall \xx,\yy \in \mathbb{R}^n$.

	\textbf{Definition~3} \textit{(Quasi-Fej\'er Convergence \cite{QF_Iusem_94}):}
	A sequence $\{ \yy^k \} \subset \mathbb{R}^n$ is quasi-Fej\'er convergent to a nonempty set $V$ if for each $\yy \in V$, there exists a non-negative integer $M$ and a sequence $\{ \epsilon^k \} \subset \mathbb{R}_{\geq 0} $ such that $\sum_{k=0}^{\infty} \epsilon_k < \infty$ and
	\vspace{-0.1cm}\begin{align}
	\norm{\yy^{k+1} - \yy}^2 \leq \norm{\yy^{k} - \yy}^2 + \epsilon_k, ~ \forall k \geq M.
	\end{align}

For the convergence analysis, we make the following assumptions:
\begin{axioms}
	\item \label{assum:coop} Considering any $\xx_j \in \Lambda_j \cap \Upsilon_j$ and $\widehat{\xx}_j \notin \Lambda_j \cap \Upsilon_j$, the inequality
    $\glkijfunc(\xx_j, \xxit{i}{n}) \leq \glkijfunc(\widehat{\xx}_j, \xxit{i}{n})$
	holds for every iteration index $n$ and $\forall \ell,k,i,j$.
	\item \label{assum:finite} The sequence of path lengths taken by the iterations of the proposed algorithms are square summable, i.e., $\sum_{n=0}^{\infty} \big( \norm{\tildexjn - \xxit{j}{n}}^2 + \norm{\xxit{j}{n+1} - \tildexjn}^2 \big) < \infty$
	for $j \in \setfromone{N_V}$.
\end{axioms}
Assumption~\ref{assum:coop} is valid especially when the cooperative algorithms can be initialized at some $\xx = (\xx_1,\ldots,\xx_{N_V})$ with $\xx_j \in \Lambda_j,\, \forall j \in \setfromone{N_V}$. Assumption~\ref{assum:coop} implies that any point inside the intersection of the noncooperative and cooperative constraint sets is closer, in terms of the function value (whose zero-level sets are the constraint sets), to the cooperative constraint sets than any point outside the intersection. When the iterations in the cooperative case start from coarse location estimates obtained in the absence of cooperation, the corresponding cooperative sets, which are dynamically changing at each iteration, may involve the set $\Lambda_j \cap \Upsilon_j$, but exclude the points outside $\Lambda_j \cap \Upsilon_j$, which yields $\glkijfunc(\xx_j, \xxit{i}{n}) \leq 0 < \glkijfunc(\widehat{\xx}_j, \xxit{i}{n})$. On the other hand, Assumption~\ref{assum:finite} represents a realistic scenario through the Armijo rule in \eqref{eq:armijo_step} and \eqref{eq:armijo_selection}, which ensures a certain level of decline in the Lambertian functions at each iteration and generates a nonincreasing sequence of step sizes.

\subsection{Quasi-Fej\'er Convergence}
In the convergence analysis, the proof of convergence is based on the concept of quasi-Fej\'er convergent sequences, which possess nice properties that facilitate further investigation, as will be presented in Lemma~2. The following proposition establishes the quasi-Fej\'er convergence of the sequences generated by Algorithm~\ref{alg:sim} to the set $\Lambda_j \cap \Upsilon_j$.

	\textbf{Proposition~1:} \textit{Assume \ref{assum:coop} and \ref{assum:finite} hold. Let $\{ \xx^{(n)} \}_{n=0}^{\infty} $ be any sequence generated by Algorithm~\ref{alg:sim}, where $\xx^{(n)} \triangleq \left( \xxit{1}{n}, \ldots, \xxit{N_V}{n} \right)$. Then, for each $j \in \setfromone{N_V}$, the sequence $\{ \xxit{j}{n} \}_{n=0}^{\infty} $ is quasi-Fej\'er convergent to the set $\Lambda_j \cap \Upsilon_j$.}

\indent\textit{Proof}: Please see Appendix~\ref{sec:sup_prop1}.

The following proposition states the quasi-Fej\'er convergence of the sequences generated by Algorithm~\ref{alg:cyclic}.

	\textbf{Proposition~2:} \textit{Assume \ref{assum:coop} and \ref{assum:finite} hold. Let $\{ \xx^{(n)} \}_{n=0}^{\infty} $ be any sequence generated by Algorithm~\ref{alg:cyclic}, where $\xx^{(n)} \triangleq \left( \xxit{1}{n}, \ldots, \xxit{N_V}{n} \right)$. Then, for each $j \in \setfromone{N_V}$, the sequence $\{ \xxit{j}{n} \}_{n=0}^{\infty} $ is quasi-Fej\'er convergent to the set $\Lambda_j \cap \Upsilon_j$.}

\indent\textit{Proof}: Please see Appendix~\ref{sec:sup_prop2}.

As the quasi-Fej\'er convergence of the sequences generated by the proposed algorithms is stated, the following lemma presents the properties of quasi-Fej\'er convergent sequences.

	\textbf{Lemma~2} \textit{(Theorem 4.1 in \cite{QF_Iusem_94}):}
	\textit{If a sequence $\{\yy^k\}$ is quasi-Fej\'er convergent to a nonempty set $V$, the following conditions hold:}
	\begin{enumerate}
		\item \textit{$\{\yy^k\}$ is bounded.}
		\item \textit{If $V$ contains an accumulation point of $\{\yy^k\}$, then $\{\yy^k\}$ converges to a point $\yy \in V$.}
	\end{enumerate}
	

\subsection{Limiting Behavior of Step Size Sequences}

In this part, we investigate the limiting behavior of the step size sequences, which are updated according to the procedure in Algorithm~\ref{alg:armijo}. The following two lemmas prove that the step size sequences generated by Algorithm~\ref{alg:sim} and Algorithm~\ref{alg:cyclic} have positive limits.

	\textbf{Lemma~3:} \textit{Any step size sequence $\lambdajn$ generated by Algorithm~\ref{alg:sim} has a positive limit, i.e.,}
	\begin{align}
	\lim_{n \to \infty} \lambdajn > 0.
	\end{align}

\indent\textit{Proof}: Please see Appendix~\ref{sec:sup_lemma3}.

\textbf{Lemma~4:} \textit{Any step size sequences $\lambda_{j,\rm{nc}}^{(n)}$ and $\lambda_{j,\rm{c}}^{(n)}$ generated by Algorithm~\ref{alg:cyclic} have positive limits, i.e.,}
\begin{equation}
\lim_{n \to \infty} \lambda_{j,\rm{nc}}^{(n)} > 0 {~~\textrm{and}~~}
\lim_{n \to \infty} \lambda_{j,\rm{c}}^{(n)} > 0.
\end{equation}

\indent\textit{Proof}: Please see Appendix~\ref{sec:sup_lemma4}.

Lemma~3 and Lemma~4 will prove to be useful for deriving the fundamental convergence properties of the proposed algorithms, as investigated next. 

\subsection{Main Convergence Results}\label{sec:main_conv}
In this part, we present the main convergence results for the proposed algorithms, i.e., convergence to a solution of Problem~1.

\textbf{Proposition~3:} \textit{Let $\{ \xx^{(n)} \}_{n=0}^{\infty} $ be any sequence generated by Algorithm~\ref{alg:sim}, where $\xx^{(n)} \triangleq \left( \xxit{1}{n}, \ldots, \xxit{N_V}{n} \right)$. Then, for each $j \in \setfromone{N_V}$, the sequence $\{ \xxit{j}{n} \}_{n=0}^{\infty} $ converges to a point $\xx_j \in \Lambda_j \cap \Upsilon_j$, i.e., a solution of Problem~1.}
\begin{proof}
	From Proposition~1, $\sum_{n=0}^{\infty} \epsilon_j^{(n)} < \infty$, where $\epsilon_j^{(n)}$ is given by \eqref{eq:eps_fejer} in Appendix~\ref{sec:sup_prop1}. Hence, $\lim_{n \to \infty} \epsilon_j^{(n)} = 0$ is obtained. Based on Lemma~3, \eqref{eq:theta_jn}, and \eqref{eq:eps_fejer}, it follows that
	\vspace{-0.2cm}\begin{align}\nonumber
	&\lim_{n \to \infty} \normbigg{\sum_{k=1}^{K_j} \Bigg( \sum_{\ell \in \tildeSkj} \tildekappa \Hflambda{\glkjfunc}(\tildexjn) \\ \label{eq:norm_limit}
    & ~ + \sum_{i=1,i \neq j}^{N_V} \sum_{\ell \in \Skij} \kappalkij \Hflambda{\glkijfunc (.,\xxit{i}{\hat{n}})}(\tildexjn) \Bigg) } = 0~,
	\end{align}
	
\vspace{-0.2cm}

\noindent which implies that
	\vspace{-0.2cm}\begin{align}\label{eq:limit_j}
	\lim_{n \to \infty} J_{f}(\tildexjn) = 0
	\end{align}
		
\vspace{-0.2cm}

\noindent is satisfied $\forall f \in \mtFtilde_j \cup \mtF_j$, where the operator $J_f$ defined on $\mathbb{R}^d$ for the set of continuously differentiable functions $f: \mathbb{R}^d \rightarrow \mathbb{R}$ is given by
	\vspace{-0.3cm}\begin{align} \label{eq:Jf}
	J_f(\xx) = \left(\frac{f^{+}(\xx)}{\norm{\nabla f(\xx)}}\right)^2 .
	\end{align}
	For a generic Lambertian function in \eqref{eq:lemma1_func}, the norm square of the gradient can be expressed as
	\vspace{-0.1cm}\begin{align}\label{eq:norm_gradient}
& \norm{\nabla g_{\epsilon}(\xx)}^2 = \frac{1}{\left(\norm{\xx-\yy}^k + \epsilon\right)^2}
\\\nonumber
& + \left(\frac{(\yy-\xx)^T \nr}{\norm{\xx-\yy}^k + \epsilon}\right)^2
\frac{ k \norm{\xx-\yy}^{k-2} \left( (k-2) \norm{\xx-\yy}^{k} - 2\epsilon \right)}{\left(\norm{\xx-\yy}^k + \epsilon\right)^2}\cdot
	\end{align}
	Since the sequence of iterates $\{\tildexjn\}_{n=0}^{\infty}$ is bounded by Lemma~2, $\big\{\norm{\tildexjn-\yy}\big\}_{n=0}^{\infty}$ is also bounded, which implies the boundedness of $\norm{\nabla g_{\epsilon}(\xx)}$. Therefore, based on \eqref{eq:limit_j} and \eqref{eq:Jf}, it follows that
	\begin{align}\label{eq:lim_func}
	\lim_{n \to \infty} f^{+}(\tildexjn) = 0, ~~ \forall f \in \mtFtilde_j \cup \mtF_j \,.
	\end{align}
	From the Bolzano-Weierstrass Theorem \cite[Section 3.4]{RealAnalysisBook}, the boundedness of $\{\tildexjn\}_{n=0}^{\infty}$ requires that $\{\tildexjn\}_{n=0}^{\infty}$ has a convergent subsequence. Denote this subsequence by $\{\tildexjnt\}_{t=0}^{\infty}$ and its limit by $\xx_j^{\star}$. From \eqref{eq:lim_func}, it turns out that $\xx_j^{\star} \in \Lambda_j \cap \Upsilon_j$. Therefore, $\Lambda_j \cap \Upsilon_j$ contains a limit point of $\{\tildexjn\}_{n=0}^{\infty}$, which, based on Lemma~2, yields the result that $\{\tildexjn\}_{n=0}^{\infty}$ converges to a point inside $\Lambda_j \cap \Upsilon_j$. Based on \eqref{eq:proj_polyhedron} and the fact that $\Lambda_j \subset \Gamma_j$, it follows that the sequence $\{\xxit{j}{n}\}_{n=0}^{\infty}$ converges to a point $\xx_j \in \Lambda_j \cap \Upsilon_j$.
\end{proof}

\textbf{Proposition~4:} \textit{Let $\{ \xx^{(n)} \}_{n=0}^{\infty} $ be any sequence generated by Algorithm~\ref{alg:cyclic}, where $\xx^{(n)} \triangleq \left( \xxit{1}{n}, \ldots, \xxit{N_V}{n} \right)$. Then, for each $j \in \setfromone{N_V}$, the sequence $\{ \xxit{j}{n} \}_{n=0}^{\infty} $ converges to a point $\xx_j \in \Lambda_j \cap \Upsilon_j$, i.e., a solution of Problem~1.}
	
\begin{proof}
Applying similar steps to those in the proof of Proposition~3 and exploiting Proposition~2 and Lemma~4, the following results are obtained:
\begin{align}\label{eq:lim_func_cyc}
\lim_{n \to \infty} \Big[ \tilde{g}_{\hat{\ell}_{\rm{nc}},\hat{k}_{\rm{nc}}}^{(j)}(\tildexjn) \Big]^{+} = 0 \\ \label{eq:lim_func_cyc_coop}
\lim_{n \to \infty} \Big[ g_{\hat{\ell}_{\rm{c}},\hat{k}_{\rm{c}}}^{(\hat{i}_{\rm{c}},j)}(\tildexjn,\ux_i^{(\hat{n})}) \Big]^{+} = 0
\end{align}
Based on the most violated constraint control in \eqref{eq:mvcc_noncoop} and \eqref{eq:mvcc_coop}, it is obvious that
\begin{align}
\tilde{f}(\tildexjn) \leq \tilde{g}_{\hat{\ell}_{\rm{nc}},\hat{k}_{\rm{nc}}}^{(j)}(\tildexjn), ~\forall \tilde{f} \in \mtFtilde_j \\
f(\tildexjn) \leq g_{\hat{\ell}_{\rm{c}},\hat{k}_{\rm{c}}}^{(\hat{i}_{\rm{c}},j)}(\tildexjn,\ux_i^{(\hat{n})}), ~\forall f \in \mtF_j
\end{align}
which implies via \eqref{eq:lim_func_cyc} and \eqref{eq:lim_func_cyc_coop} that
\begin{align}\label{eq:lim_func_cyc2}
\lim_{n \to \infty} f^{+}(\tildexjn) = 0, ~~ \forall f \in \mtFtilde_j \cup \mtF_j ~.
\end{align}
The rest of the proof is the same as that in Proposition~3.
\end{proof}



\vspace{-0.2cm}

\section{Numerical Results}\label{sec:nume}

In this section, numerical examples are provided to investigate the theoretical bounds on cooperative localization in VLP networks and to evaluate the performance of the proposed projection-based algorithms. The VLP network parameters are determined in a similar manner to the work in \cite{CRB_TOA_VLC} and \cite{MFK_CRLB}. The area of each PD is set to $1\,$cm$^2$ and the Lambertian order of all the LEDs is selected as $m=1$. In addition, the noise variances are calculated using \cite[Eq.~6]{Lampe_VLC_TCOM_2015}. The parameters for noise variance calculation are set to be the same as those used in \cite{Lampe_VLC_TCOM_2015} (see Table~I in \cite{Lampe_VLC_TCOM_2015}).

\begin{figure}
	\center
	\vspace{-0.1cm}
	\includegraphics[scale=0.55]{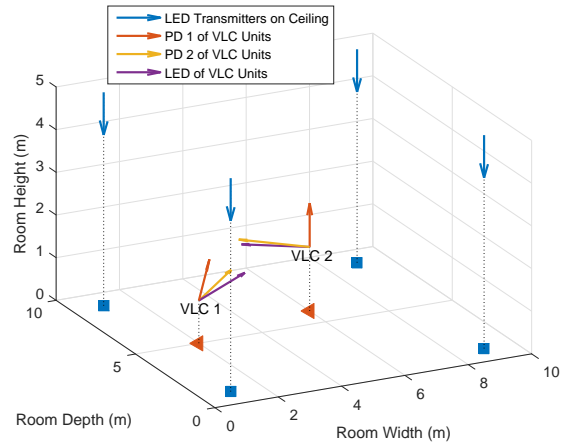}
	\vspace{-0.4cm}
	\caption{VLP network configuration in the simulations. Each VLC unit contains two PDs and one LED. PD~$1$ of the VLC units is used to obtain measurements from the LEDs on the ceiling while PD~$2$ of the VLC units communicates with the LED of the other VLC unit for cooperative localization. The squares and the triangles show the projections of the LEDs and the VLC units on the floor, respectively.}\label{fig:VLP_config_3d}
	\vspace{-0.6cm}
\end{figure}

The VLP network considered in the simulations is illustrated in Fig.~\ref{fig:VLP_config_3d}.
A room of size $10$m$\times 10$m$\times 5$m is considered, where there exist $N_L = 4$ LED transmitters on the ceiling which are located at $\uy_1 = \left[1~1~5\right]^T$m, $\uy_2 = \left[1~9~5\right]^T$m, $\uy_3 = \left[9~1~5\right]^T$m, and $\uy_4 = \left[9~9~5\right]^T$m. The LEDs on the ceiling have perpendicular orientations, i.e., $\tildenTj = \left[ 0 ~~  0  ~ -1 \right]^T$ for $j \in \{1,2,3,4\}$. In addition, there exist $N_V = 2$ VLC units whose locations are given by $\ux_1 = \left[ 2~5~1 \right]^T$m and $\ux_2 = \left[ 6~6~1.5 \right]^T$m. Each VLC unit consists of two PDs and one LED, with offsets with respect to the center of the VLC unit being set to $\ua_{j,1} = \left[0~-0.1~0\right]^T$m, $\ua_{j,2} = \left[0~0.1~0\right]^T$m, and $\ub_{j,1} = \left[ 0.1~0~0 \right]^T$m for $j = 1,2$. The orientation vectors of the PDs and the LEDs on the VLC units are obtained as the normalized versions (the orientation vectors are unit-norm) of the following vectors: $\un^{(1)}_{R,1} = \left[ 0.3~-0.1~1 \right]^T$, $\un^{(2)}_{R,1} = \left[ 0.2~0.4~1 \right]^T$, $\un^{(1)}_{R,2} = \left[ 0.8~0.6~0.1 \right]^T$, $\un^{(2)}_{R,2} = \left[ -0.7~0.2~0.1 \right]^T$, $\un^{(1)}_{T,1} = \left[ 0.9~0.4~0.1 \right]^T$, and $\un^{(2)}_{T,1} = \left[ -0.8~0.1~0.1 \right]^T$. Furthermore, the connectivity sets are defined as $\Skijgen{1}{i}{j} = \emptyset$, $\Skijgen{2}{i}{j} = \{1\}$ for $i, j \in \{1,2\}, i \neq j$ for the cooperative measurements and $\tildeSkjgen{1}{1} = \{1,2,3\}$, $\tildeSkjgen{1}{2} = \{2,3,4\}$ and $\tildeSkjgen{2}{j} = \emptyset$ for $j\in \{1,2\}$ for the noncooperative measurements.

\subsection{Theoretical Bounds}\label{sec:nume_crlb}

In this part, the CRLB expression derived in Section~\ref{sec:systemModel} is investigated to illustrate the effects of cooperation on the localization performance of VLP networks.


\subsubsection{{Performance with Respect to Transmit Power of LEDs on Ceiling}}
\begin{figure}[t]
	\center
	\vspace{-0.2cm}
	\includegraphics[scale=0.55]{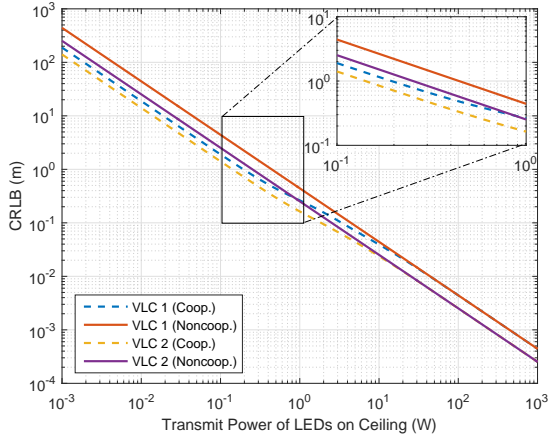}
	\vspace{-0.2cm}
	\caption{Individual CRLBs for localization of VLC units in both noncooperative and cooperative cases with respect to the transmit power of LEDs on ceiling, where the transmit power of VLC units is taken as $1$W.}\label{fig:individual_CRLB_vs_LED_power}
	\vspace{-0.5cm}
\end{figure}

In order to analyze the localization performance of the VLC units with respect to the transmit powers of
the LEDs on the ceiling (equivalently, anchors), individual CRLBs for localization of the VLC units in noncooperative and cooperative scenarios are plotted against the transmit powers of LEDs on the ceiling in Fig.~\ref{fig:individual_CRLB_vs_LED_power}, where the transmit powers of the VLC units are fixed to $1\,$W. As observed from Fig.~\ref{fig:individual_CRLB_vs_LED_power}, {cooperation among VLC units can provide substantial improvements in localization accuracy (about $81\,$cm and $37\,$cm improvement, respectively, for VLC~$1$ and VLC~$2$ for the LED transmit power of $300\,$mW).} {We note that} the improvement gained by employing cooperation is higher for VLC~$1$ as compared to that for VLC~$2$. This is an intuitive result since the localization of VLC~$1$ depends mostly on LED~$1$ {and LED~$2$} (the other LEDs are not sufficiently close to facilitate the localization process), and incorporating cooperative measurements for VLC~$1$ provides an enhancement in localization performance that is much greater than that for VLC $2$, which can obtain informative measurements from the LEDs on the ceiling even in the absence of cooperation {as seen from the network geometry in Fig.~\ref{fig:VLP_config_3d}. In addition,} the CRLBs in the cooperative scenario converge to those in the noncooperative scenario as the transmit powers of the LEDs increase. Since the first (second) summand in the FIM expression in \eqref{eq:FIM2} corresponds to the noncooperative (cooperative) localization, higher transmit powers of the LEDs on the ceiling cause the first summand to be much greater than the second summand, which makes the contribution of cooperation to the FIM negligible. Hence, the effect of cooperation on localization performance becomes {more} significant as the transmit power {decreases}, which is in compliance with the results obtained for RF based cooperative localization networks \cite{BookChapter_Coop_2016}.

\subsubsection{{Performance with Respect to Transmit Power of VLC Units}}
\begin{figure}[t]
	\center
	\vspace{-0.2cm}
	\includegraphics[scale=0.55]{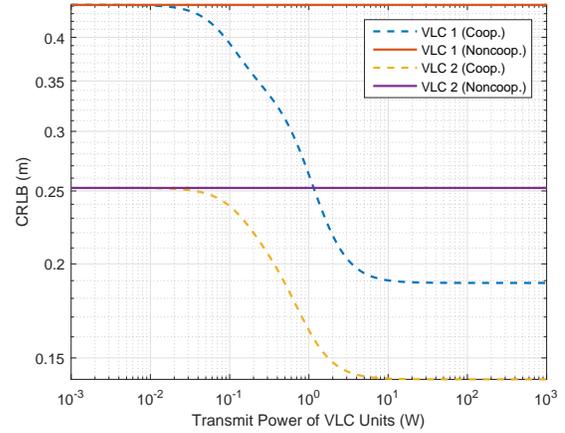}
	\vspace{-0.2cm}
	\caption{Individual CRLBs for localization of VLC units in both noncooperative and cooperative cases with respect to the transmit power of VLC units, where the transmit power of LEDs on ceiling is taken as $1$W.}\label{fig:individual_CRLB_vs_VLC_power}
	\vspace{-0.5cm}
\end{figure}

Secondly, the localization performance of the VLC units is investigated with respect to the transmit powers of the VLC units when the transmit powers of the LEDs on the ceiling are fixed to $1\,$W. Fig.~\ref{fig:individual_CRLB_vs_VLC_power} illustrates the CRLBs for localization of the VLC units against the transmit powers of the VLC units in the noncooperative and cooperative cases. As observed from Fig.~\ref{fig:individual_CRLB_vs_VLC_power}, cooperation leads to a higher improvement in the performance of VLC~$1$, similar to Fig.~\ref{fig:individual_CRLB_vs_LED_power}. In addition, via the FIM expression in \eqref{eq:FIM2}, it can be noted that the contribution of cooperation to localization performance gets higher as the transmit powers of the VLC units increase, which is also observed from Fig.~\ref{fig:individual_CRLB_vs_VLC_power}. However, the CRLB reaches a saturation level above a certain power threshold, as opposed to Fig.~\ref{fig:individual_CRLB_vs_LED_power}, where the CRLB continues to decrease as the power increases. The main reason for this distinction between the effects of the transmit powers of the LEDs on the ceiling and those of the VLC units can be explained as follows: For a fixed transmit power of the VLC units, the localization error by using {three} anchors (i.e., {three} LEDs on the ceiling {that are connected to the corresponding VLC unit}) converges to zero as the transmit powers of the anchors increase regardless of the existence of cooperation. On the other hand, for a fixed transmit power of the LEDs on the ceiling, increasing the transmit power of the VLC unit (i.e., one of the anchors) cannot reduce the localization error below a certain level. Therefore, the saturation level represents the localization accuracy that can be attained by {four} anchors with {three} anchors leading to noisy RSS measurements and one anchor generating noise-free RSS measurements.

\vspace{-0.2cm}

\subsection{Performance of the Proposed Algorithms}\label{sec:nume_alg}

In this part, the proposed algorithms in Algorithm~\ref{alg:cyclic} (CCGP) and Algorithm~\ref{alg:sim} (CSGP) are evaluated in terms of localization performance and convergence speed. For both algorithms, the initial step size is selected as $\lambda_0 = 1$, the step size shrinkage factor and the degree of decline in the Armijo rule in Algorithm~\ref{alg:armijo} are set to $\xi = 0.5$ and $\beta = 0.001$, respectively. The VLC units are initialized at the positions of the closest LEDs on the ceiling which are connected to the corresponding VLC units.

Localization performances of the algorithms are presented in both the absence and the presence of cooperation and compared against those of the ML estimator in \eqref{eq:MLE} and the CRLBs derived in Section~\ref{sec:systemModel}. In order to ensure convergence to the global minimum, the ML estimator is implemented using a multi-start optimization algorithm with $100$ initial points randomly selected from the interval $[0~ 10]\,$m at each axis.\footnote{The implemented estimator is effectively a maximum \textit{a posteriori} probability (MAP) estimator with a uniform prior distribution over the interval $[0~ 10]\,$m, based on the prior information that VLC units are inside the room. {Hence, the implemented ML estimator may achieve smaller RMSEs than the CRLB in the low SNR regime, where the prior information becomes more significant as the measurements are very noisy.}} In addition, two different measurement noise distributions, namely, Gaussian and exponential, are considered while evaluating the proposed algorithms as in \cite{Gholami_2011_Eurasip_CFP}. The Gaussian noise is used to model the case in which the RSS measurement noise can be both positive and negative, whereas the exponentially distributed noise (subtracted from the true value) represents the scenario in which the RSS measurements are negatively biased, which leads to the feasibility modeling of the localization problem in Section~\ref{sec:problem_formulation}. Furthermore, the average residuals at each iteration are calculated to assess the convergence speed of the proposed algorithms \cite{Gholami_ICFP_TSP_2013}:
\vspace{-0.2cm}\begin{equation}\label{eq:avg_residual}
\varrho_n = \frac{1}{{M N_V}}\sum_{m=1}^{M} \norm{ \xx^{(n,m)} - \xx^{(n-1,m)} }
\end{equation}

\vspace{-0.2cm}

\noindent where $\xx^{(n,m)} = \big(\xx_{1}^{(n,m)}, \ldots, \xx_{N_V}^{(n,m)} \big)$ denotes the position vector of all the VLC units at the $n$th iteration for the $m$th Monte Carlo realization of measurement noises and $M$ is the number of Monte Carlo realizations.

In the simulations, two-dimensional localization is performed by assuming that the VLC units have known heights. Therefore, with the knowledge of perpendicular LED orientations, Case~2 type Lambertian sets in Section~\ref{sec:known_height_perp_LED} are utilized for localization based on the measurements from the LEDs on the ceiling. The cooperation among the VLC units is modeled by Case~1 type Lambertian sets in Section~\ref{sec:expanded_Lambertian}.

%

%
%

\subsubsection{{Gaussian Noise}}
\begin{figure}
	\centering
	\vspace{-0.3cm}
	\includegraphics[scale=0.55]{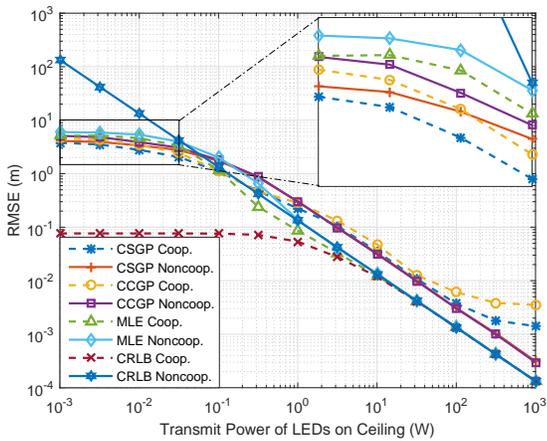}
	\vspace{-0.2cm}
	\caption{Average localization error of VLC units with respect to the transmit power of LEDs on ceiling for the proposed algorithms in Algorithm~\ref{alg:cyclic} (CCGP) and Algorithm~\ref{alg:sim} (CSGP) along with the MLE and CRLB for the case of Gaussian measurement noise.}\label{fig:rmse_pos_neg_noise}
	\vspace{-0.5cm}
\end{figure}

In Fig.~\ref{fig:rmse_pos_neg_noise}, the average localization errors of the VLC units for the different algorithms are plotted against the transmit power of the LEDs on the ceiling for the case of the Gaussian measurement noise by fixing the transmit powers of the LEDs at the VLC units to $1\,$W. From Fig.~\ref{fig:rmse_pos_neg_noise}, it is observed that the cooperative approach can significantly reduce the localization errors, especially in the low SNR regime ({about $60~\rm{cm}$ and $70~\rm{cm}$ reduction} for CSGP and CCGP algorithms, respectively, for $100~\rm{mW}$ LED transmit power). In addition, both Algorithm~\ref{alg:cyclic} (CCGP) and Algorithm~\ref{alg:sim} (CSGP) can attain the localization error levels that asymptotically converge to zero at the same rate as that of the CRLB. Moreover, it can be inferred from Fig.~\ref{fig:rmse_pos_neg_noise} that the proposed iterative methods achieve higher localization performance than the ML estimator in the low SNR regime for both the noncooperative and the cooperative scenarios. Although the ML estimator is forced to converge to the global minimum via the multi-start optimization procedure involving $100$ different executions of a convex optimization solver, whose complexity may be prohibitive for practical implementations, it has lower performance than the proposed approaches, which depend on low-complexity iterative gradient projections. Hence, at low SNRs, the proposed algorithms are superior to the MLE in terms of both the localization performance and the computational complexity. Furthermore, the simultaneous projections outperforms the cyclic projections at low SNRs at the cost of a higher number of set projections, but the two approaches converge asymptotically as the SNR increases.


%

\begin{figure}
	\begin{center}
		\vspace{-0.3cm}
		\subfigure[]{
			\label{fig:conv_speed_gauss_100mW}
			\includegraphics[scale=0.55]{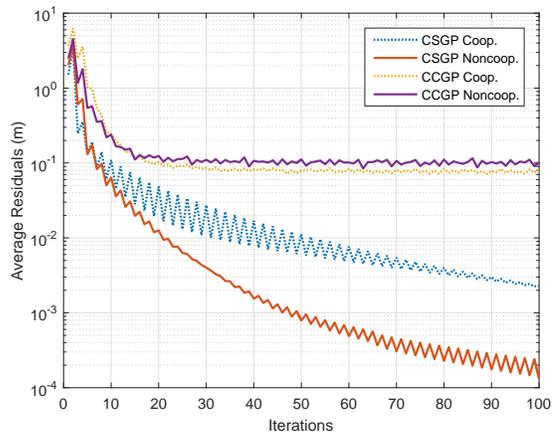}
		}
		\vspace{-0.3cm}
		\subfigure[]{
			\label{fig:conv_speed_gauss_1W}
			\includegraphics[scale=0.55]{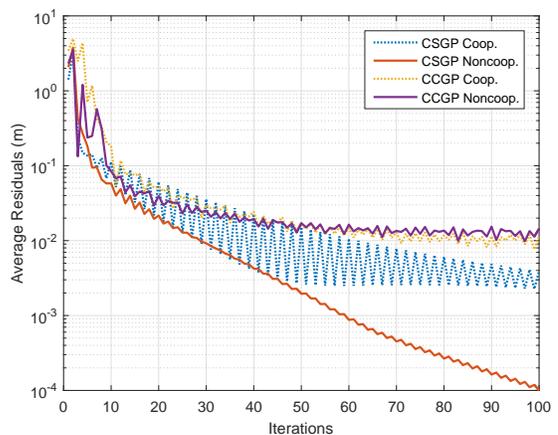}
		}
	\end{center}
	\caption{Convergence rate of the average residuals in \eqref{eq:avg_residual} for the proposed algorithms in Algorithm~\ref{alg:cyclic} and Algorithm~\ref{alg:sim} for the case of Gaussian measurement noise, where the transmit power of LEDs on ceiling is \subref{fig:conv_speed_gauss_100mW} $100\,$mW and \subref{fig:conv_speed_gauss_1W} $1\,$W.}
	\label{fig:conv_speed_gauss}
	\vspace{-0.5cm}
\end{figure}

Fig.~\ref{fig:conv_speed_gauss_100mW} and Fig.~\ref{fig:conv_speed_gauss_1W} report the average residuals calculated by \eqref{eq:avg_residual} corresponding to the proposed algorithms versus the number of iterations for $100\,$mW and $1\,$W of transmit powers of the LEDs on the ceiling, respectively. CSGP in the absence of cooperation has the fastest convergence rate and exhibits an almost monotonic convergence behavior. However, CSGP in the cooperative scenario shows relatively slow convergence in general and a locally nonmonotonic behavior when several consecutive iterations are taken into account. This is due to the cooperative Lambertian sets being involved in the simultaneous projection operations. In addition, cyclic projections tend to settle into limit cycle oscillations after few iterations, thus implying that the sequence itself does not converge to a point, but it has several subsequences that converge \cite{Censor_Subgradient_2008}. This behavior, called \textit{cyclic convergence}, is encountered in cyclic (sequential) projections if the feasibility problem is inconsistent \cite{Censor_Subgradient_2008,Gubin1967}. Furthermore, by comparing Fig.~\ref{fig:conv_speed_gauss_100mW} and Fig.~\ref{fig:conv_speed_gauss_1W}, it is observed that the magnitude of limit cycle oscillations gets smaller as the SNR increases since the region of uncertainty becomes narrower at higher SNR values, thereby making the convergent subsequences close to each other.

\subsubsection{{Exponential Noise}}
\begin{figure}
	\centering
	\vspace{-0.1cm}
	\includegraphics[scale=0.55]{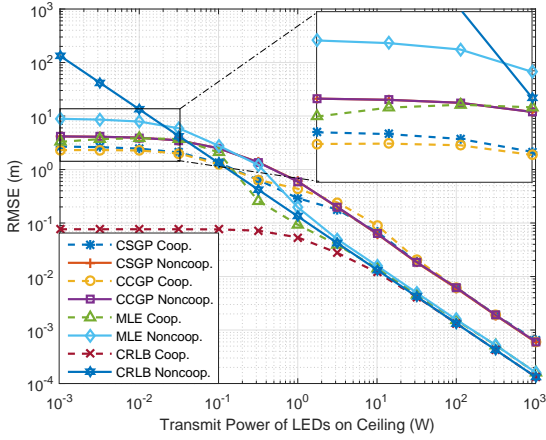}
	\vspace{-0.3cm}
	\caption{Average localization error of VLC units with respect to the transmit power of LEDs on ceiling for the proposed algorithms in Algorithm~\ref{alg:cyclic} (CCGP) and Algorithm~\ref{alg:sim} (CSGP) along with the MLE and CRLB for the case of exponentially distributed measurement noise.}\label{fig:AvgLocError_vs_LEDPower_conn_3_exp}
	\vspace{-0.4cm}
\end{figure}

To investigate the performance of the algorithms under exponentially distributed measurement noise, the average localization errors are plotted against the transmit power of the LEDs on the ceiling for the case of the subtractive exponential noise in Fig.~\ref{fig:AvgLocError_vs_LEDPower_conn_3_exp}. Similar to the case of the Gaussian noise, the proposed algorithms succeed in converging to the true VLC unit positions as the SNR increases. Since the projection based methods rely on the assumption of negatively biased measurements, they perform slightly better at low SNRs as compared to the case of the Gaussian noise. On the other hand, the MLE produces larger errors at low SNRs for the exponentially distributed noise since its derivation is based on the assumption of Gaussian noise.

\begin{figure}
	\begin{center}
		\subfigure[]{
			\label{fig:conv_speed_gauss_100mW_exp}
			\includegraphics[scale=0.55]{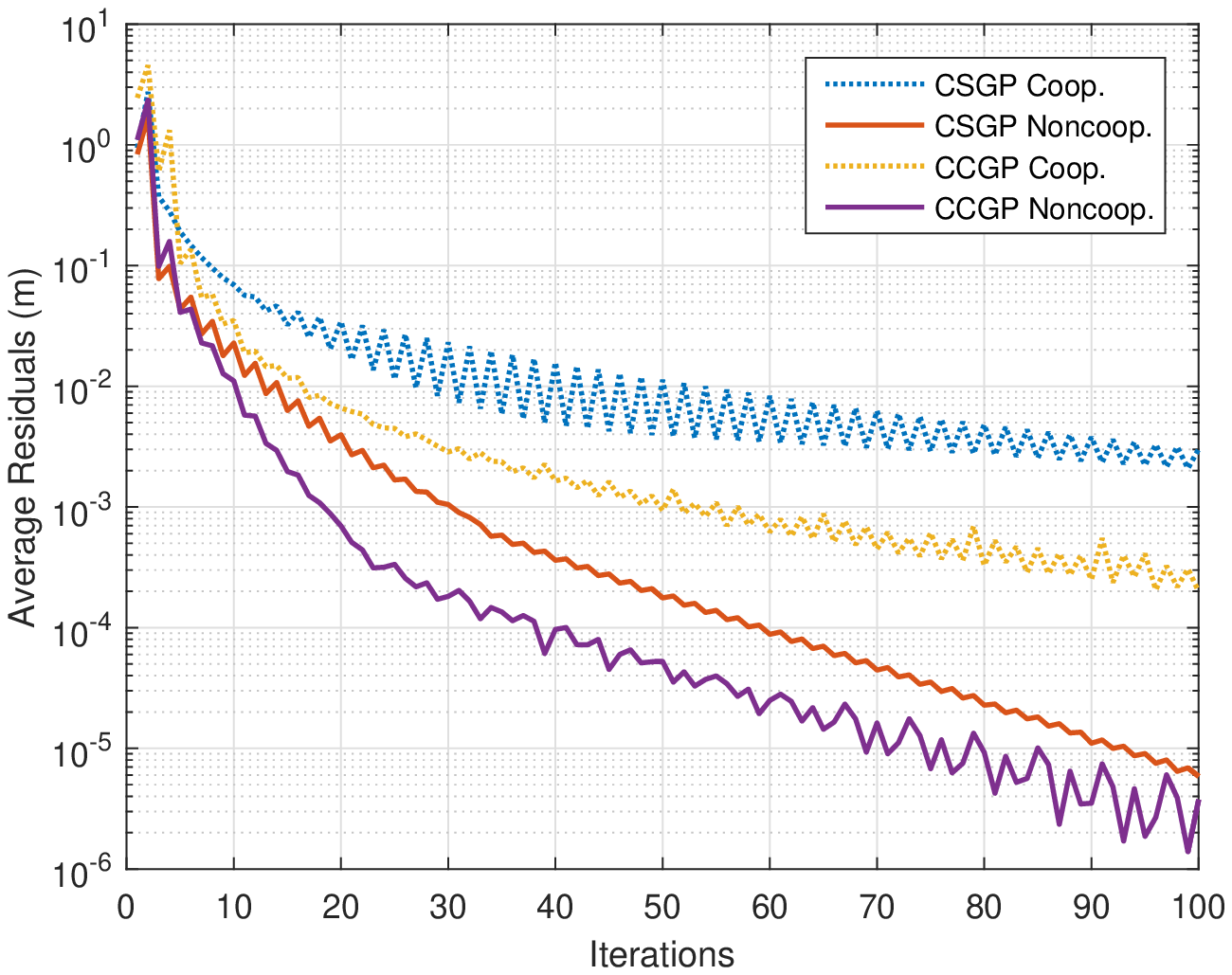}
		}
		\vspace{-0.3cm}
		\subfigure[]{
			\label{fig:conv_speed_gauss_1W_exp}
			\includegraphics[scale=0.55]{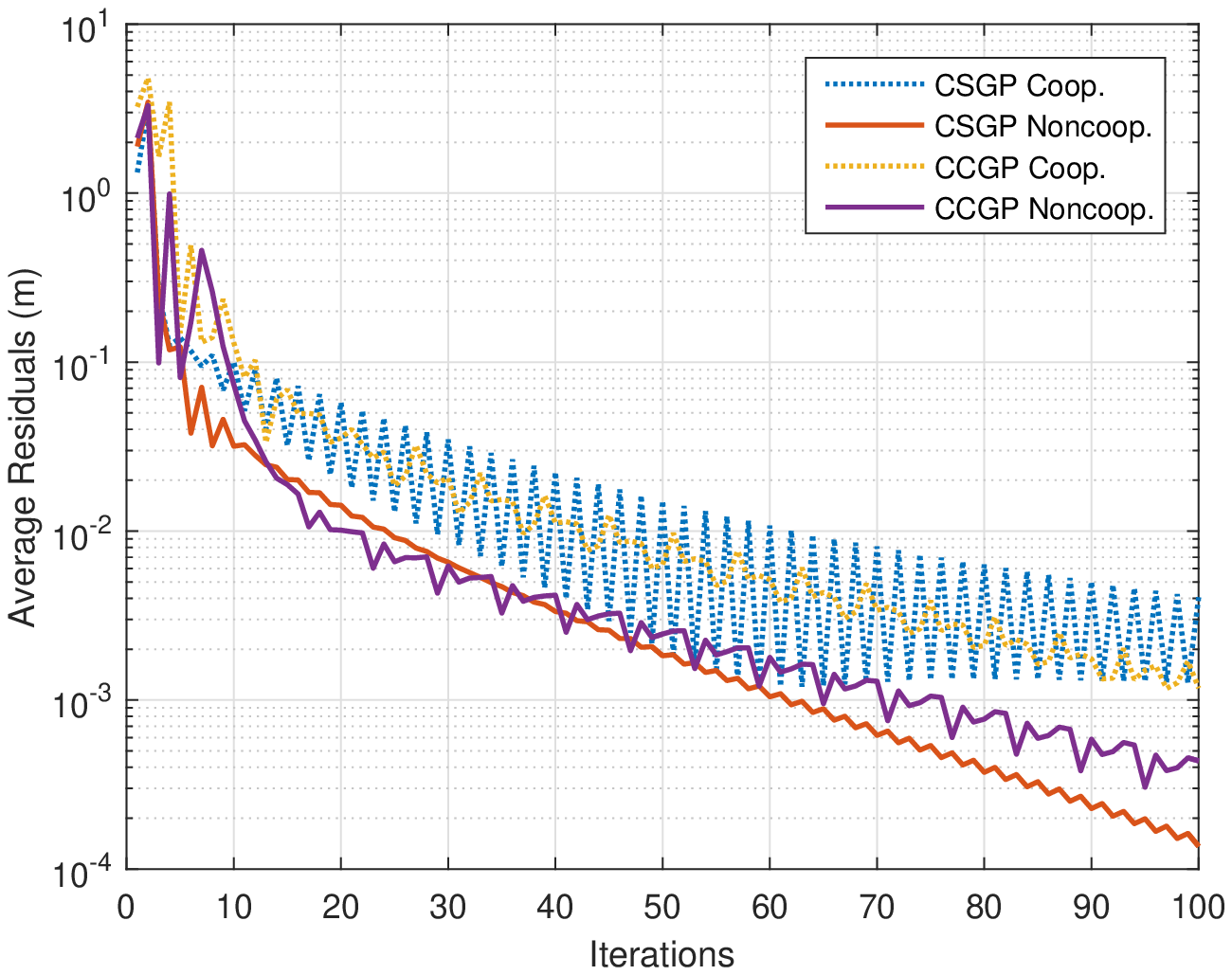}
		}
	\end{center}
	\caption{Convergence rate of the average residuals in \eqref{eq:avg_residual} for the proposed algorithms in Algorithm~\ref{alg:cyclic} and Algorithm~\ref{alg:sim} for the case of exponentially distributed measurement noise, where the transmit power of LEDs on ceiling is \subref{fig:conv_speed_gauss_100mW_exp} $100\,$mW and \subref{fig:conv_speed_gauss_1W_exp} $1\,$W.}
	\label{fig:conv_speed_gauss_1W_exp_overall}
\end{figure}

The average residuals in the case of the exponentially distributed noise are illustrated in Fig.~\ref{fig:conv_speed_gauss_100mW_exp} and Fig.~\ref{fig:conv_speed_gauss_1W_exp} for two different LED power levels. In contrary to the case of Gaussian noise, cyclic projections do not fall into limit cycles and provide globally monotonic convergence results as the feasibility problem is consistent, which complies with the results presented in the literature pertaining to the study of CFPs \cite{Censor_Subgradient_2008}. In addition, it is observed that both the cyclic and the sequential projection methods have faster convergence for lower SNR values since it takes fewer iterations to get inside the intersection of the constraint sets, which becomes larger as the SNR decreases.

\section{Concluding Remarks}\label{sec:conc}

In this manuscript, a cooperative VLP network has been proposed based on a generic system model consisting of LED transmitters at known locations and VLC units with multiple LEDs and PDs. First, the CRLB on the overall localization error of the VLC units has been derived to quantify the effects of cooperation on the localization accuracy of VLP networks. Then, due to the nonconvex nature of the corresponding ML {expression}, the problem of cooperative localization has been formulated as {a QFP}, which facilitates the development of low-complexity decentralized feasibility-seeking methods. In order to solve the feasibility problem, iterative gradient projections based algorithms have been proposed. Furthermore, based on the notion of quasi-Fej\'er convergent sequences, formal convergence proofs have been provided for the proposed algorithms in the consistent case. Finally, numerical examples {have been} presented to illustrate the significance of cooperation in VLP networks and to investigate the performance of the proposed algorithms in terms of localization accuracy and convergence speed. It has been verified that the proposed iterative methods {asymptotically} converge to the true positions of VLC units at high SNR and exhibit superior performance over the ML estimator {at low SNRs} in terms of both implementation complexity and localization accuracy.

An important research direction for future studies is to explore the convergence properties of Algorithm~\ref{alg:cyclic} and Algorithm~\ref{alg:sim} when the proposed QFP is inconsistent. In the inconsistent case, simultaneous projection algorithms tend to converge to a minimizer of a \textit{proximity function} that specifies the distance to constraint sets \cite{Censor_Steered_Inconsistent_2004,Censor_Subgradient_2008}. For the implicit CFP (ICFP) considered in TOA-based wireless network localization, the POCS based simultaneous algorithm is shown to converge to the minimizer of a convex function, which is the sum of squares of the distances to the constraint sets \cite{Gholami_ICFP_TSP_2013}. Therefore, finding proximity functions characterizing the behavior of simultaneous projections {(e.g., Algorithm~\ref{alg:sim})} for the inconsistent QFPs \cite{Censor_QFP_2006} would be a significant extension for the set-theoretic estimation literature.

\vspace{-0.2cm}

\appendix

\subsection{Proof of Lemma~1}\label{sec:sup_lemma1}

	Suppose that $\xx_1 \in \mtB$, $\xx_2 \in \mtB$, and $\alpha < \gamma$. It is clear that for any $\lambda \in \left(0,1\right)$, $\lambda \xx_1 + (1-\lambda) \xx_2 \in \Omega$. Also, for any $\lambda \in \left(0,1\right)$,
\begin{align}\label{lemma_11}
&  g_{\epsilon}(\lambda \xx_1 + (1-\lambda) \xx_2)  \\
& = \gamma - \frac{\lambda(\yy-\xx_1)^T \nr + (1-\lambda)(\yy-\xx_2)^T \nr}{\norm{\lambda(\xx_1-\yy) + (1-\lambda)(\xx_2-\yy)}^k + \epsilon} \\ \label{lemma_12}
& \leq \gamma - \frac{\lambda(\yy-\xx_1)^T \nr + (1-\lambda)(\yy-\xx_2)^T \nr}{\lambda\norm{\xx_1-\yy}^k + (1-\lambda)\norm{\xx_2-\yy}^k + \epsilon} \\
& \leq \gamma - \nonumber \\  \label{lemma_14}
& \frac{\lambda (\gamma-\alpha) (\norm{\xx_1-\yy}^k + \epsilon) + (1-\lambda) (\gamma-\alpha) (\norm{\xx_2-\yy}^k + \epsilon)}{\lambda(\norm{\xx_1-\yy}^k + \epsilon) + (1-\lambda)(\norm{\xx_2-\yy}^k + \epsilon)} \\ \label{lemma_15}
& = \alpha
\end{align}
is obtained, where \eqref{lemma_12} is due to the convexity of $\norm{.}^k$, $\xx_1 \in \Omega$, and $\xx_2 \in \Omega$, and \eqref{lemma_14} follows from $\xx_1 \in \mtB$ and $\xx_2 \in \mtB$. Hence, \eqref{lemma_11}--\eqref{lemma_15} implies the convexity of $\mtB$ for $\alpha < \gamma$. For $\alpha \geq \gamma$, $g_{\epsilon}(\xx) \leq \alpha$ is satisfied $\forall \xx \in \Omega$, which implies $\mtB = \Omega$. Therefore, $\mtB$ is convex $\forall \alpha \in \mathbb{R}$.

\subsection{Proof of Proposition~1}\label{sec:sup_prop1}

Since $\Lambda_j \cap \Upsilon_j \neq \emptyset$, consider any point $\xx_j \in \Lambda_j \cap \Upsilon_j$. At the $n$th iteration, it can be assumed that $\xxit{j}{n} \notin \Lambda_j \cap \Upsilon_j$ because otherwise iterations will stop via \eqref{eq:sim_proj} and \eqref{eq:gradient_projector}, which implies quasi-Fej\'er convergence of $\{ \xxit{j}{n} \}_{n=0}^{\infty} $ to $\Lambda_j \cap \Upsilon_j$ based on Definition~3. Then, based on the iterative step in \eqref{eq:sim_proj}, the following is obtained:
\begin{align}\label{eq:norm_eq}
\norm{\xxit{j}{n+1} - \xx_j}^2 =
\norm{ \tilde{\xx}_{j}^{(n)} - \xx_j - \lambda_j^{(n)} \xthetajn }^2
\end{align}
where
\vspace{-0.3cm}\begin{align}\nonumber
\xthetajn &\triangleq \sum_{k=1}^{K_j} \Bigg( \sum_{\ell \in \tildeSkj} \tildekappa \Hflambda{\glkjfunc}(\tildexjn) \\ \label{eq:theta_jn} &+ \sum_{i=1,i \neq j}^{N_V} \sum_{\ell \in \Skij} \kappalkij \Hflambda{\glkijfunc (.,\xxit{i}{\hat{n}})}(\tildexjn) \Bigg)
\end{align}
with the scaled gradient operator being defined as
\vspace{-0.1cm}\begin{align}\label{eq:scaled_gradient}
\Hflambda{f}(\xx) = \frac{f^{+}(\xx)}{\norm{\nabla f(\xx)}^2} \nabla f(\xx).
\end{align}

\vspace{-0.1cm}

\noindent From \eqref{eq:norm_eq}, it follows that
\vspace{-0.1cm}\begin{align}\nonumber
\norm{\xxit{j}{n+1} - \xx_j}^2 &= \norm{\tildexjn - \xx_j}^2 + \left(\lambdajn\right)^2 \norm{\xthetajn}^2 \\ \label{eq:norm_3}
&- 2 \lambdajn \left(\xthetajn\right)^T \left( \tildexjn - \xx_j \right).
\end{align}

\vspace{-0.2cm}

\noindent Let $\mtFtilde_j$ and $\mtF_j$ be the sets of functions for the $j$th VLC unit corresponding to the noncooperative and cooperative cases, respectively, which are given by
\vspace{-0.1cm}\begin{align} \label{eq:Ftildej}
&\mtFtilde_j = \left\{ \{ \glkjfunc \}_{\ell \in \tildeSkj} \right\}_{k \in \setfromone{K_j}}
\\\label{eq:Fj}
&\mtF_j = \left\{ \{ \{ \glkijfunc(.,\xxit{i}{\hat{n}}) \}_{\ell \in \tildeSkj} \}_{i \in \setfromone{N_V} \setminus j} \right\}_{k \in \setfromone{K_j}}
\end{align}

\vspace{-0.2cm}

\noindent For any function $f \in \mtFtilde_j \cup \mtF_j$, $f(\xx_j) \leq 0$ holds since $\xx_j \in \Lambda_j \cap \Upsilon_j$ (see \eqref{eq:Lambertian_set_noncoop}, \eqref{eq:Lambertian_set_coop}, \eqref{eq:set_all_noncoop}, and \eqref{eq:set_all_coop}). Consider the following two subsets of $\mtFtilde_j \cup \mtF_j$:
$\mtFbreve_{j,n}^{\star} = \{ f \in \mtFtilde_j \cup \mtF_j ~|~ f(\tildexjn) \leq 0 \}$ and
$\mtFbreve_{j,n}^{\diamond} = \{ f \in \mtFtilde_j \cup \mtF_j ~|~ f(\tildexjn) > 0 \}$.
It is clear from \eqref{eq:scaled_gradient} that for any $f^{\star} \in \mtFbreve_{j,n}^{\star}$
\vspace{-0.2cm}\begin{align}\label{eq:f0}
\Hflambda{f^{\star}}(\tildexjn) = 0
\end{align}

\vspace{-0.2cm}

\noindent is satisfied. On the other hand, for any $f^{\diamond} \in \mtFbreve_{j,n}^{\diamond} \cap \mtFtilde_j$, $f^{\diamond}(\xx_j) = 0 < f^{\diamond}(\tildexjn)$, and, for any $f^{\diamond} \in \mtFbreve_{j,n}^{\diamond} \cap \mtF_j$, $f^{\diamond}(\xx_j) < f^{\diamond}(\tildexjn)$ via Assumption~\ref{assum:coop}. Then, the following inequality holds for any $f^{\diamond} \in \mtFbreve_{j,n}^{\diamond}$:
\vspace{-0.2cm}\begin{align}\label{eq:f1}
f^{\diamond}(\xx_j) < f^{\diamond}\big(\tildexjn\big)~.
\end{align}

\vspace{-0.2cm}

\noindent Since $\xx_j$ and $\tildexjn$ both lie inside the halfspaces of the form \eqref{eq:halfspace} and \eqref{eq:halfspace_coop} corresponding to the set of functions $\mtFbreve_{j,n}^{\diamond}$ ($\xx_j \in \Lambda_j \subset \Gamma_j$ and $\tildexjn \in \Gamma_j$, see \eqref{eq:polyhedron}, \eqref{eq:halfspace}, \eqref{eq:proj_polyhedron} and \eqref{eq:set_all_noncoop}), they are in the region where any $f^{\diamond} \in \mtFbreve_{j,n}^{\diamond}$ is quasiconvex. Hence, from \eqref{eq:f1} and Definition~2,
$\big(\nabla f^{\diamond}(\tildexjn)\big)^T \big( \xx_j - \tildexjn \big)\leq 0$
follows, which, based on \eqref{eq:scaled_gradient}, implies that
\begin{align}\label{eq:result_quasiconvex_2}
\left( \Hflambda{f^{\diamond}}(\tildexjn) \right)^T \left( \tildexjn - \xx_j \right) \geq 0 ~.
\end{align}
The inner product term in \eqref{eq:norm_3} can be decomposed into two parts corresponding to the sets $\mtFbreve_{j,n}^{\star}$ and $\mtFbreve_{j,n}^{\diamond}$. The part that corresponds to $\mtFbreve_{j,n}^{\star}$ is $0$ via \eqref{eq:f0} and the remaining part is greater than or equal to $0$ via \eqref{eq:result_quasiconvex_2}. Hence, the following inequality is obtained:
$\big(\xthetajn\big)^T \big( \tildexjn - \xx_j \big) \geq 0$,
which, based on \eqref{eq:norm_3}, yields
\vspace{-0.1cm}\begin{align} \label{eq:fejer_1}
\norm{\xxit{j}{n+1} - \xx_j}^2 \leq \norm{\tildexjn - \xx_j}^2 + \epsilon_j^{(n)}
\end{align}
where
\vspace{-0.3cm}\begin{align} \label{eq:eps_fejer}
\epsilon_j^{(n)} \triangleq \left(\lambdajn\right)^2 \norm{\xthetajn}^2 .
\end{align}
From the fact that $\xx_j \in \Gamma_j$, the following can be written:
\vspace{-0.1cm}\begin{align} \label{eq:norm_proj1}
\norm{\tildexjn - \xx_j} &= \norm{P_{\Gamma_j}(\xxit{j}{n}) - P_{\Gamma_j}(\xx_j)} \\ \label{eq:norm_proj2}
& \leq \norm{\xxit{j}{n} - \xx_j}
\end{align}
where \eqref{eq:norm_proj1} follows from \eqref{eq:proj_polyhedron}, and \eqref{eq:norm_proj2} is due to the nonexpansivity of the orthogonal projection operator. Combining \eqref{eq:norm_proj1} and \eqref{eq:norm_proj2} with \eqref{eq:fejer_1} yields the following inequality:
\begin{align} \label{eq:quasi_fejer_step}
\norm{\xxit{j}{n+1} - \xx_j}^2 \leq \norm{\xxit{j}{n} - \xx_j}^2 + \epsilon_j^{(n)} .
\end{align}
Based on the parallel projection step in \eqref{eq:sim_proj}, it can easily be shown that
\begin{align}\label{eq:series_finite2}
\norm{\xxit{j}{n+1} - \tildexjn} = \lambdajn \norm{\xthetajn}
\end{align}
where $\xthetajn$ is given by \eqref{eq:theta_jn}. Then, from Assumption~\ref{assum:finite}, it follows that $\sum_{n=0}^{\infty} \norm{\xxit{j}{n+1} - \tildexjn}^2 < \infty$, which leads to $\sum_{n=0}^{\infty} \epsilon_j^{(n)} < \infty$ via \eqref{eq:series_finite2} and \eqref{eq:eps_fejer}. Finally, using \eqref{eq:quasi_fejer_step} and Definition~3 yields the desired result.

\subsection{Proof of Proposition~2}\label{sec:sup_prop2}

Following the same steps as stated in the proof of Proposition~1, the following inequality is obtained based on \eqref{eq:cyclic_iterations}:
\begin{align} \label{eq:fejer_1_prop2}
\norm{\xxit{j}{n+1} - \xx_j}^2 \leq \norm{\xxit{j}{n} - \xx_j}^2 + \epsilon_j^{(n)}
\end{align}
where
\begin{align} \label{eq:fejer_epsilon}
\epsilon_j^{(n)} \triangleq \norm{ \vartheta_{\rm{nc}} \lambda_{j,\rm{nc}}^{(n)} \Hflambda{\tilde{g}_{\hat{\ell}_{\rm{nc}},\hat{k}_{\rm{nc}}}^{(j)}} (\tildexjn) + \vartheta_{\rm{c}} \lambda_{j,\rm{c}}^{(n)} \Hflambda{g_{\hat{\ell}_{\rm{c}},\hat{k}_{\rm{c}}(.,\xxit{i}{\hat{n}})}^{(\hat{i}_{\rm{c}},j)}} (\tildexjn) }^2
\end{align}
with $\Hflambda{f}$ being defined as in \eqref{eq:scaled_gradient}. The averaging step in \eqref{eq:cyclic_iterations} leads to
$\norm{\xxit{j}{n+1} - \tildexjn} = \sqrt{\epsilon_j^{(n)}}$,
where $\epsilon_j^{(n)}$ is given by \eqref{eq:fejer_epsilon}. Assuming that \ref{assum:finite} holds and following an approach similar to that in the proof of Proposition~1, the inequality $\sum_{n=0}^{\infty} \epsilon_j^{(n)} < \infty$ is obtained, thus establishing the quasi-Fej\'er convergence of the sequence $\{ \xxit{j}{n} \}_{n=0}^{\infty} $ to $\Lambda_j \cap \Upsilon_j$.

\subsection{Proof of Lemma~3}\label{sec:sup_lemma3}

The proof is based on contradiction. Suppose that $\lim_{n \to \infty} \lambdajn = 0$. Then, for each $\zeta > 0$, there exists an iteration index $n(\zeta)$ such that $\lambda_j^{(n(\zeta))} < \zeta$. Based on the Armijo step size selection rule \eqref{eq:armijo_selection} in Algorithm~\ref{alg:armijo} and the step size update equation \eqref{eq:alg_sim_step_size} in Algorithm~\ref{alg:sim}, there exists a function $f^{\circ} \in \mtFtilde_j \cup \mtF_j$ such that the inequality in \eqref{eq:armijo_selection} is not satisfied for the step size
$\tilde{\zeta} = \lambda_j^{(n(\zeta)-1)} \xi^{\bar{m}}$,
where
$\bar{m} = \max\{ m \in \mathbb{Z}_{\geq 0} ~ | ~ \lambda_j^{(n(\zeta)-1)} \xi^{m} > \zeta \}$.
Hence, the following inequality is obtained:
\begin{align}\label{eq:step_size_ineq}
f^{\circ} ( \Gflambda{f^{\circ}}{\tilde{\zeta}} (\tildexjnzeta) ) > \fcirc(\tildexjnzeta)(1 - \beta \tildezeta)
\end{align}
It is clear that $\fcirc(\tildexjnzeta) > 0$ since otherwise the step size selection procedure would not need to be applied, meaning that $\tildexjnzeta$ is inside the zero-level set of every function in $\mtFtilde_j \cup \mtF_j$, i.e., $\tildexjnzeta \in \Lambda_j \cap \Upsilon_j$, which completes the proof of convergence of the iterates $\{ \xxit{j}{n} \}_{n=0}^{\infty} $ to the set $\Lambda_j$ via Lemma~2. Then, the left hand side of \eqref{eq:step_size_ineq} can be rewritten using the Taylor series expansion and \eqref{eq:gradient_projector2} as follows:
\begin{align}\label{eq:taylor}
\fcirc ( \Gflambda{\fcirc}{\tildezeta} (\tildexjnzeta) ) = \fcirc(\tildexjnzeta) - \tildezeta \fcirc(\tildexjnzeta) + O(\tildezeta^2)
\end{align}
where $O(\tildezeta^2)$ represents the terms with $\tildezeta^s$ for $s \geq 2$. Since $\lim_{\tildezeta \to 0} O(\tildezeta^2)/\tildezeta = 0$, there exists $\upsilon > 0$ such that
\begin{align}\label{eq:exist_ups}
{O(\tildezeta^2)}/{\tildezeta} < \fcirc(\tildexjnzeta)(1-\beta)
\end{align}
is satisfied for $0 < \tildezeta \leq \upsilon$. The existence of $\upsilon$ satisfying \eqref{eq:exist_ups} is guaranteed by $\fcirc(\tildexjnzeta) > 0$ and $\beta \in (0,1)$. Inserting \eqref{eq:exist_ups} into \eqref{eq:taylor} yields the following inequality:
$\fcirc ( \Gflambda{\fcirc}{\tildezeta} (\tildexjnzeta) ) < \fcirc (\tildexjnzeta) (1 - \beta \tildezeta)$,
which contradicts with the inequality in \eqref{eq:step_size_ineq}. Therefore, the initial assumption is not valid, which implies $\lim_{n \to \infty} \lambdajn > 0$.

\subsection{Proof of Lemma~4}\label{sec:sup_lemma4}

	The proof can be obtained by invoking similar arguments to those in the proof of Lemma~3 and using the step size update rule in \eqref{eq:step_size_cyc_nc} and \eqref{eq:step_size_cyc_c}.

\subsection{Partial Derivatives in \eqref{eq:FIM2}}\label{sec:sup_partial}
	
\noindent From \eqref{eq:alp1} and \eqref{eq:alp2}, the partial derivatives in \eqref{eq:FIM2} are obtained as follows:
\begin{align}\nonumber
&\frac{\partial\widetilde{\alpha}_{l,k}^{(j)}(\ux_j)}{\partial x_{t}}=
-\frac{(\tildeml + 1)\tildePTl \Akj\big((\utildedlkj)^{T} \tilden_{T,l} \big)^{\tildeml}}{2\pi\norm{\utildedlkj} ^{\tildeml + 3}}
\\\nonumber
&\times\Big(
\tildeml\widetilde{n}_{T,l}(t-3j+3)(\utildedlkj)^T \nRkj\big((\utildedlkj)^{T} \tilden_{T,l} \big)^{-1}
\\\label{eq:partialAlpTil}
&\quad~+{n}_{R,k}^{(j)}(t-3j+3)
\\\nonumber
&\quad~-(\tildeml+3)\widetilde{d}^{(j)}_{l,k}(t-3j+3)(\utildedlkj)^T \nRkj\norm{\utildedlkj}^{-2}
\Big)
\end{align}
for $t\in\{3j-2,3j-1,3j\}$ and ${\partial\widetilde{\alpha}_{l,k}^{(j)}(\ux_j)}/{\partial x_{t}}=0$ otherwise, where $\widetilde{n}_{T,l}(t-3j+3)$, ${n}_{R,k}^{(j)}(t-3j+3)$, and $\widetilde{d}^{(j)}_{l,k}(t-3j+3)$ represent the $(t-3j+3)$th elements of $\tilden_{T,l}$, $\nRkj$, and $\utildedlkj$, respectively. Similarly,
\begin{align}\nonumber
&\frac{\partial{\alpha}_{l,k}^{(i,j)}(\ux_j,\ux_i)}{\partial x_{t}}=
-\frac{(\mli + 1) \PTli \Akj \big( (\udlkij)^{T} \nTli \big)^{\mli}}
{2\pi\norm{\udlkij}^{\mli + 3}}
\\\nonumber
&\times\Big(
\mli n_{T,l}^{(i)}(t-3j+3)(\udlkij)^T \nRkj\big( (\udlkij)^{T} \nTli \big)^{-1}
\\\label{eq:partialAlp}
&\quad~+n_{R,k}^{(j)}(t-3j+3)
\\\nonumber
&\quad~-(\mli+3)d^{(i,j)}_{l,k}(t-3j+3) (\udlkij)^T\nRkj \norm{\udlkij}^{-2}
\Big)
\end{align}
for $t\in\{3j-2,3j-1,3j\}$, ${\partial{\alpha}_{l,k}^{(i,j)}(\ux_i,\ux_j)}/{\partial x_{t}}$ is equal to the negative of \eqref{eq:partialAlp} with $(t-3j+3)$'s being replaced by $(t-3i+3)$'s for $t\in\{3i-2,3i-1,3i\}$, and ${\partial{\alpha}_{l,k}^{(i,j)}(\ux_i,\ux_j)}/{\partial x_{t}}=0$ otherwise. In \eqref{eq:partialAlp}, $n_{T,l}^{(i)}(t-3j+3)$ and $d^{(i,j)}_{l,k}(t-3j+3)$ denote the $(t-3j+3)$th elements of $\nTli$ and $\udlkij$, respectively.

\bibliographystyle{IEEEtran}
\bibliography{references2,Ref4}

\begin{thebibliography}{10}
\providecommand{\url}[1]{#1}
\csname url@samestyle\endcsname
\providecommand{\newblock}{\relax}
\providecommand{\bibinfo}[2]{#2}
\providecommand{\BIBentrySTDinterwordspacing}{\spaceskip=0pt\relax}
\providecommand{\BIBentryALTinterwordstretchfactor}{4}
\providecommand{\BIBentryALTinterwordspacing}{\spaceskip=\fontdimen2\font plus
\BIBentryALTinterwordstretchfactor\fontdimen3\font minus
  \fontdimen4\font\relax}
\providecommand{\BIBforeignlanguage}[2]{{%
\expandafter\ifx\csname l@#1\endcsname\relax
\typeout{** WARNING: IEEEtran.bst: No hyphenation pattern has been}%
\typeout{** loaded for the language `#1'. Using the pattern for}%
\typeout{** the default language instead.}%
\else
\language=\csname l@#1\endcsname
\fi
#2}}
\providecommand{\BIBdecl}{\relax}
\BIBdecl

\bibitem{WSNloc_book}
G.~Mao and B.~Fidan, \emph{Localization Algorithms and Strategies for Wireless
  Sensor Networks}.\hskip 1em plus 0.5em minus 0.4em\relax Information Science
  Reference, 2009.

\bibitem{pahlavan2013principles}
K.~Pahlavan and P.~Krishnamurthy, \emph{Principles of Wireless Access and
  Localization}, ser. Wiley Desktop Editions.\hskip 1em plus 0.5em minus
  0.4em\relax Wiley, 2013.

\bibitem{HLIU}
H.~Liu, H.~Darabi, P.~Banerjee, and J.~Liu, ``Survey of wireless indoor
  positioning techniques and systems,'' \emph{IEEE Transactions on Systems,
  Man, and Cybernetics, Part C (Applications and Reviews)}, vol.~37, no.~6, pp.
  1067--1080, Nov. 2007.

\bibitem{Sinan_Survey}
S.~Gezici, ``A survey on wireless position estimation,'' \emph{Wireless
  Personal Communications}, vol.~44, no.~3, pp. 263--282, Feb. 2008.

\bibitem{HandbookPos}
R.~Zekavat and R.~M. Buehrer, \emph{Handbook of Position Location: Theory,
  Practice and Advances}.\hskip 1em plus 0.5em minus 0.4em\relax John Wiley \&
  Sons, 2011.

\bibitem{bookSahin}
Z.~Sahinoglu, S.~Gezici, and I.~Guvenc, \emph{Ultra-wideband Positioning
  Systems: Theoretical Limits, Ranging Algorithms, and Protocols}.\hskip 1em
  plus 0.5em minus 0.4em\relax New York: Cambridge University Press, 2008.

\bibitem{VLP_Roadmap}
J.~Armstrong, Y.~Sekercioglu, and A.~Neild, ``Visible light positioning: {A}
  roadmap for international standardization,'' \emph{IEEE Communications
  Magazine}, vol.~51, no.~12, pp. 68--73, Dec. 2013.

\bibitem{SurveyVLC15}
P.~H. Pathak, X.~Feng, P.~Hu, and P.~Mohapatra, ``Visible light communication,
  networking, and sensing: A survey, potential and challenges,'' \emph{IEEE
  Communications Surveys Tutorials}, vol.~17, no.~4, pp. 2047--2077,
  Fourthquarter 2015.

\bibitem{BeyondPoint}
H.~Burchardt, N.~Serafimovski, D.~Tsonev, S.~Videv, and H.~Haas, ``{VLC:
  Beyond} point-to-point communication,'' \emph{IEEE Communications Magazine},
  vol.~52, no.~7, pp. 98--105, July 2014.

\bibitem{VLP_CRLB_RSS}
X.~Zhang, J.~Duan, Y.~Fu, and A.~Shi, ``Theoretical accuracy analysis of indoor
  visible light communication positioning system based on received signal
  strength indicator,'' \emph{Journal of Lightwave Technology}, vol.~32,
  no.~21, pp. 4180--4186, Nov. 2014.

\bibitem{zhang2014asynchronous}
W.~Zhang, M.~I.~S. Chowdhury, and M.~Kavehrad, ``Asynchronous indoor
  positioning system based on visible light communications,'' \emph{Optical
  Engineering}, vol.~53, no.~4, pp. {045\,105--1}--{045\,105--9}, 2014.

\bibitem{CRB_TOA_VLC}
T.~Wang, Y.~Sekercioglu, A.~Neild, and J.~Armstrong, ``Position accuracy of
  time-of-arrival based ranging using visible light with application in indoor
  localization systems,'' \emph{Journal of Lightwave Technology}, vol.~31,
  no.~20, pp. 3302--3308, Oct. 2013.

\bibitem{MFK_CRLB}
M.~F. Keskin and S.~Gezici, ``Comparative theoretical analysis of distance
  estimation in visible light positioning systems,'' \emph{Journal of Lightwave
  Technology}, vol.~34, no.~3, pp. 854--865, Feb. 2016.

\bibitem{TDOA_VLC}
S.-Y. Jung, S.~Hann, and C.-S. Park, ``{TDOA-based} optical wireless indoor
  localization using {LED} ceiling lamps,'' \emph{IEEE Transactions on Consumer
  Electronics}, vol.~57, no.~4, pp. 1592--1597, Nov. 2011.

\bibitem{GuvencWAMI15}
Y.~Eroglu, I.~Guvenc, N.~Pala, and M.~Yuksel, ``{AOA-based} localization and
  tracking in multi-element {VLC} systems,'' in \emph{IEEE 16th Annual Wireless
  and Microwave Technology Conference (WAMICON),}, Apr. 2015.

\bibitem{RSS_aperture_JLT_2017}
H.~Steendam, T.~Q. Wang, and J.~Armstrong, ``Theoretical lower bound for indoor
  visible light positioning using received signal strength measurements and an
  aperture-based receiver,'' \emph{Journal of Lightwave Technology}, vol.~35,
  no.~2, pp. 309--319, Jan 2017.

\bibitem{ZZB_MFK}
M.~F. Keskin, E.~Gonendik, and S.~Gezici, ``Improved lower bounds for ranging
  in synchronous visible light positioning systems,'' \emph{Journal of
  Lightwave Technology}, vol.~34, no.~23, pp. 5496--5504, Dec 2016.

\bibitem{LocatingNodes_Patwari_2005}
N.~Patwari, J.~N. Ash, S.~Kyperountas, A.~O. Hero, R.~L. Moses, and N.~S.
  Correal, ``Locating the nodes: cooperative localization in wireless sensor
  networks,'' \emph{IEEE Signal Processing Magazine}, vol.~22, no.~4, pp.
  54--69, July 2005.

\bibitem{Henk_Cooperative}
H.~Wymeersch, J.~Lien, and M.~Z. Win, ``Cooperative localization in wireless
  networks,'' \emph{Proc. IEEE}, vol.~97, no.~2, pp. 427--450, Feb. 2009.

\bibitem{BookChapter_Coop_2016}
M.~R. Gholami, M.~F. Keskin, S.~Gezici, and M.~Jansson, \emph{Cooperative
  Positioning in Wireless Networks}.\hskip 1em plus 0.5em minus 0.4em\relax
  Wiley Encyclopedia of Electrical and Electronics Engineering, 2016, pp.
  1--19.

\bibitem{SDP_Biswas_2006}
P.~Biswas, T.-C. Lian, T.-C. Wang, and Y.~Ye, ``Semidefinite programming based
  algorithms for sensor network localization,'' \emph{{ACM} Trans. Sens.
  Netw.}, vol.~2, no.~2, pp. 188--220, 2006.

\bibitem{SDP_Chan_2009_TSP}
K.~W.~K. Lui, W.~K. Ma, H.~C. So, and F.~K.~W. Chan, ``Semi-definite
  programming algorithms for sensor network node localization with
  uncertainties in anchor positions and/or propagation speed,'' \emph{IEEE
  Transactions on Signal Processing}, vol.~57, no.~2, pp. 752--763, Feb 2009.

\bibitem{RSS_SDP_TVT_2010}
R.~W. Ouyang, A.~K.~S. Wong, and C.~T. Lea, ``Received signal strength-based
  wireless localization via semidefinite programming: Noncooperative and
  cooperative schemes,'' \emph{IEEE Transactions on Vehicular Technology},
  vol.~59, no.~3, pp. 1307--1318, March 2010.

\bibitem{SOCP_Tseng_2007}
P.~Tseng, ``Second-order cone programming relaxation of sensor network
  localization,'' \emph{{SIAM} J. Optim.}, vol.~18, no.~1, pp. 156--185, Feb.
  2007.

\bibitem{SOCP_Lampe_TWCOM_2014}
G.~Naddafzadeh-Shirazi, M.~B. Shenouda, and L.~Lampe, ``Second order cone
  programming for sensor network localization with anchor position
  uncertainty,'' \emph{IEEE Transactions on Wireless Communications}, vol.~13,
  no.~2, pp. 749--763, February 2014.

\bibitem{Soares_Relax_TSP_2015}
C.~Soares, J.~Xavier, and J.~Gomes, ``Simple and fast convex relaxation method
  for cooperative localization in sensor networks using range measurements,''
  \emph{IEEE Transactions on Signal Processing}, vol.~63, no.~17, pp.
  4532--4543, Sept 2015.

\bibitem{Blatt_POCS_TSP_2006}
D.~Blatt and A.~O. Hero, ``Energy-based sensor network source localization via
  projection onto convex sets,'' \emph{IEEE Transactions on Signal Processing},
  vol.~54, no.~9, pp. 3614--3619, Sept 2006.

\bibitem{Gholami_2011_Eurasip_CFP}
\BIBentryALTinterwordspacing
M.~R. Gholami, H.~Wymeersch, E.~G. Str{\"o}m, and M.~Rydstr{\"o}m, ``Wireless
  network positioning as a convex feasibility problem,'' \emph{EURASIP Journal
  on Wireless Communications and Networking}, vol. 2011, no.~1, p. 161, 2011.
  [Online]. Available: \url{http://dx.doi.org/10.1186/1687-1499-2011-161}
\BIBentrySTDinterwordspacing

\bibitem{Gholami_ICFP_TSP_2013}
M.~R. Gholami, L.~Tetruashvili, E.~G. Str\"om, and Y.~Censor, ``Cooperative
  wireless sensor network positioning via implicit convex feasibility,''
  \emph{IEEE Transactions on Signal Processing}, vol.~61, no.~23, pp.
  5830--5840, Dec 2013.

\bibitem{DistProj_TWCOM_2015}
Y.~Zhang, Y.~Lou, Y.~Hong, and L.~Xie, ``Distributed projection-based
  algorithms for source localization in wireless sensor networks,'' \emph{IEEE
  Transactions on Wireless Communications}, vol.~14, no.~6, pp. 3131--3142,
  June 2015.

\bibitem{MDS_2006}
\BIBentryALTinterwordspacing
J.~A. Costa, N.~Patwari, and A.~O. Hero, III, ``Distributed
  weighted-multidimensional scaling for node localization in sensor networks,''
  \emph{ACM Trans. Sen. Netw.}, vol.~2, no.~1, pp. 39--64, Feb. 2006. [Online].
  Available: \url{http://doi.acm.org/10.1145/1138127.1138129}
\BIBentrySTDinterwordspacing

\bibitem{Censor_QFP_2006}
\BIBentryALTinterwordspacing
Y.~Censor and A.~Segal, ``Algorithms for the quasiconvex feasibility problem,''
  \emph{Journal of Computational and Applied Mathematics}, vol. 185, no.~1, pp.
  34 -- 50, 2006. [Online]. Available:
  \url{http://www.sciencedirect.com/science/article/pii/S037704270500049X}
\BIBentrySTDinterwordspacing

\bibitem{QF_Iusem_94}
\BIBentryALTinterwordspacing
A.~N. Iusem, B.~F. Svaiter, and M.~Teboulle, ``Entropy-like proximal methods in
  convex programming,'' \emph{Mathematics of Operations Research}, vol.~19,
  no.~4, pp. 790--814, 1994. [Online]. Available:
  \url{http://www.jstor.org/stable/3690314}
\BIBentrySTDinterwordspacing

\bibitem{SetJia_TMC_2011}
T.~Jia and R.~M. Buehrer, ``A set-theoretic approach to collaborative position
  location for wireless networks,'' \emph{IEEE Transactions on Mobile
  Computing}, vol.~10, no.~9, pp. 1264--1275, 2011.

\bibitem{Censor_CFP_SSP_2012}
\BIBentryALTinterwordspacing
A.~Carmi, Y.~Censor, and P.~Gurfil, ``Convex feasibility modeling and
  projection methods for sparse signal recovery,'' \emph{Journal of
  Computational and Applied Mathematics}, vol. 236, no.~17, pp. 4318 -- 4335,
  2012. [Online]. Available:
  \url{//www.sciencedirect.com/science/article/pii/S0377042712001410}
\BIBentrySTDinterwordspacing

\bibitem{Combettes_TIP_97}
P.~L. Combettes, ``Convex set theoretic image recovery by extrapolated
  iterations of parallel subgradient projections,'' \emph{IEEE Transactions on
  Image Processing}, vol.~6, no.~4, pp. 493--506, Apr 1997.

\bibitem{ICFP_Censor_Denoising_2016}
Y.~Censor, A.~Gibali, F.~Lenzen, and C.~Schnorr, ``The implicit convex
  feasibility problem and its application to adaptive image denoising,''
  \emph{arXiv:1606.05848}, 2016.

\bibitem{censor_survey_2010}
Y.~Censor and A.~Segal, ``Iterative projection methods in biomedical inverse
  problems,'' in \emph{Procceding of the interdisciplinary workshop on
  Mathematical Methods in Biomedical Imaging and Intensity-Modulated Radiation
  Therapy (IMRT)}, Pisa, Italy, Oct. 2008, pp. 65--96.

\bibitem{Censor_Subgradient_2008}
\BIBentryALTinterwordspacing
D.~Butnariu, Y.~Censor, P.~Gurfil, and E.~Hadar, ``On the behavior of
  subgradient projections methods for convex feasibility problems in
  {E}uclidean spaces,'' \emph{SIAM Journal on Optimization}, vol.~19, no.~2,
  pp. 786--807, 2008. [Online]. Available:
  \url{http://dx.doi.org/10.1137/070689127}
\BIBentrySTDinterwordspacing

\bibitem{Guvenc_hybrid}
A.~Sahin, Y.~S. Eroglu, I.~Guvenc, N.~Pala, and M.~Yuksel, ``Hybrid 3-{D}
  localization for visible light communication systems,'' \emph{Journal of
  Lightwave Technology}, vol.~33, no.~22, pp. 4589--4599, Nov 2015.

\bibitem{VLP_Accelerometer}
M.~Yasir, S.~W. Ho, and B.~N. Vellambi, ``Indoor positioning system using
  visible light and accelerometer,'' \emph{Journal of Lightwave Technology},
  vol.~32, no.~19, pp. 3306--3316, Oct. 2014.

\bibitem{EPSILON}
L.~Li, P.~Hu, C.~Peng, G.~Shen, and F.~Zhao, ``Epsilon: {A} visible light based
  positioning system,'' in \emph{11th USENIX Symposium on Networked Systems
  Design and Implementation}, Seattle, WA, 2014, pp. 331--343.

\bibitem{Poor}
H.~V. Poor, \emph{An Introduction to Signal Detection and Estimation}.\hskip
  1em plus 0.5em minus 0.4em\relax New York: Springer-Verlag, 1994.

\bibitem{Censor_SparsityFeas_2017}
S.~{Penfold}, R.~{Zalas}, M.~{Casiraghi}, M.~{Brooke}, Y.~{Censor}, and
  R.~{Schulte}, ``{Sparsity constrained split feasibility for dose-volume
  constraints in inverse planning of intensity-modulated photon or proton
  therapy},'' \emph{Physics in Medicine and Biology}, vol.~62, p. 3599, May
  2017.

\bibitem{Tight_OA_Letter_2016}
S.~Yousefi, H.~Wymeersch, X.~W. Chang, and B.~Champagne, ``Tight
  two-dimensional outer-approximations of feasible sets in wireless sensor
  networks,'' \emph{IEEE Communications Letters}, vol.~20, no.~3, pp. 570--573,
  March 2016.

\bibitem{Gholami_PlaneProjection_2010}
M.~R. Gholami, M.~Rydstrom, and E.~G. Strom, ``Positioning of node using plane
  projection onto convex sets,'' in \emph{2010 IEEE Wireless Communication and
  Networking Conference}, April 2010, pp. 1--5.

\bibitem{Gholami_PIMRC_2010}
M.~R. Gholami, S.~Gezici, E.~G. Str\"om, and M.~Rydstr\"om, ``A distributed
  positioning algorithm for cooperative active and passive sensors,'' in
  \emph{Proc. IEEE International Symposium on Personal, Indoor and Mobile Radio
  Communications (PIMRC)}, Sep. 2010.

\bibitem{Bauschke_Proj_ConvFeas_1996}
\BIBentryALTinterwordspacing
H.~H. Bauschke and J.~M. Borwein, ``On projection algorithms for solving convex
  feasibility problems,'' \emph{SIAM Review}, vol.~38, no.~3, pp. 367--426,
  1996. [Online]. Available: \url{http://dx.doi.org/10.1137/S0036144593251710}
\BIBentrySTDinterwordspacing

\bibitem{Mathlouthi2016}
Y.~Mathlouthi, A.~Mitiche, and I.~B. Ayed, \emph{Boundary Preserving
  Variational Image Differentiation}.\hskip 1em plus 0.5em minus 0.4em\relax
  Cham: Springer International Publishing, 2016, pp. 355--364.

\bibitem{NP_Book_2006}
M.~Bazarra, H.~Sherali, and C.~Shetty, \emph{Nonlinear Programming: Theory and
  Algorithms}.\hskip 1em plus 0.5em minus 0.4em\relax Wiley-Interscience, 2006.

\bibitem{LED_MultiRec}
S.-H. Yang, E.-M. Jung, and S.-K. Han, ``Indoor location estimation based on
  {LED} visible light communication using multiple optical receivers,''
  \emph{IEEE Communications Letters}, vol.~17, no.~9, pp. 1834--1837, Sep.
  2013.

\bibitem{Censor1982_CSP}
\BIBentryALTinterwordspacing
Y.~Censor and A.~Lent, ``Cyclic subgradient projections,'' \emph{Mathematical
  Programming}, vol.~24, no.~1, pp. 233--235, 1982. [Online]. Available:
  \url{http://dx.doi.org/10.1007/BF01585107}
\BIBentrySTDinterwordspacing

\bibitem{QuasiFejer_Combettes_2001}
\BIBentryALTinterwordspacing
P.~L. Combettes, ``Quasi-{F}ej\'erian analysis of some optimization
  algorithms,'' in \emph{Inherently Parallel Algorithms in Feasibility and
  Optimization and their Applications}, ser. Studies in Computational
  Mathematics, Y.~C. Dan~Butnariu and S.~Reich, Eds.\hskip 1em plus 0.5em minus
  0.4em\relax Elsevier, 2001, vol.~8, pp. 115 -- 152. [Online]. Available:
  \url{http://www.sciencedirect.com/science/article/pii/S1570579X01800100}
\BIBentrySTDinterwordspacing

\bibitem{vonNeumann_MAP}
J.~Von~Neumann, \emph{Functional Operators (AM-22), Volume 2: The Geometry of
  Orthogonal Spaces.(AM-22)}.\hskip 1em plus 0.5em minus 0.4em\relax Princeton
  University Press, 2016, vol.~2.

\bibitem{Halperin_1962_Product}
I.~Halperin, ``The product of projection operators,'' \emph{Acta Sci.
  Math.(Szeged)}, vol.~23, no.~1, pp. 96--99, 1962.

\bibitem{Armijo_1966}
\BIBentryALTinterwordspacing
L.~Armijo, ``Minimization of functions having {L}ipschitz continuous first
  partial derivatives,'' \emph{Pacific J. Math.}, vol.~16, no.~1, pp. 1--3,
  1966. [Online]. Available:
  \url{http://projecteuclid.org/euclid.pjm/1102995080}
\BIBentrySTDinterwordspacing

\bibitem{Wolfe_1969}
\BIBentryALTinterwordspacing
P.~Wolfe, ``Convergence conditions for ascent methods,'' \emph{SIAM Review},
  vol.~11, no.~2, pp. 226--235, 1969. [Online]. Available:
  \url{http://www.jstor.org/stable/2028111}
\BIBentrySTDinterwordspacing

\bibitem{nonlin_program_Bertsekas_99}
D.~P. Bertsekas, \emph{Nonlinear Programming}, 2nd~ed.\hskip 1em plus 0.5em
  minus 0.4em\relax Athena Scientific, 1999.

\bibitem{Fliege2000}
\BIBentryALTinterwordspacing
J.~Fliege and B.~F. Svaiter, ``Steepest descent methods for multicriteria
  optimization,'' \emph{Mathematical Methods of Operations Research}, vol.~51,
  no.~3, pp. 479--494, 2000. [Online]. Available:
  \url{http://dx.doi.org/10.1007/s001860000043}
\BIBentrySTDinterwordspacing

\bibitem{DosSantos1987}
\BIBentryALTinterwordspacing
L.~T.~D. Santos, ``A parallel subgradient projections method for the convex
  feasibility problem,'' \emph{Journal of Computational and Applied
  Mathematics}, vol.~18, no.~3, pp. 307 -- 320, 1987. [Online]. Available:
  \url{//www.sciencedirect.com/science/article/pii/0377042787900045}
\BIBentrySTDinterwordspacing

\bibitem{boyd2006randomized}
S.~Boyd, A.~Ghosh, B.~Prabhakar, and D.~Shah, ``Randomized gossip algorithms,''
  \emph{IEEE/ACM Transactions on Networking (TON)}, vol.~14, no.~SI, pp.
  2508--2530, 2006.

\bibitem{mangasarian1994nonlinear}
O.~L. Mangasarian, \emph{Nonlinear programming}.\hskip 1em plus 0.5em minus
  0.4em\relax SIAM, 1994.

\bibitem{RealAnalysisBook}
R.~G. Bartle and D.~R. Sherbert, \emph{Introduction to Real Analysis}.\hskip
  1em plus 0.5em minus 0.4em\relax John Wiley \& Sons, 2000.

\bibitem{Lampe_VLC_TCOM_2015}
H.~Ma, L.~Lampe, and S.~Hranilovic, ``Coordinated broadcasting for multiuser
  indoor visible light communication systems,'' \emph{IEEE Transactions on
  Communications}, vol.~63, no.~9, pp. 3313--3324, Sep. 2015.

\bibitem{Gubin1967}
\BIBentryALTinterwordspacing
L.~Gubin, B.~Polyak, and E.~Raik, ``The method of projections for finding the
  common point of convex sets,'' \emph{\{USSR\} Computational Mathematics and
  Mathematical Physics}, vol.~7, no.~6, pp. 1 -- 24, 1967. [Online]. Available:
  \url{//www.sciencedirect.com/science/article/pii/0041555367901139}
\BIBentrySTDinterwordspacing

\bibitem{Censor_Steered_Inconsistent_2004}
Y.~Censor, A.~R.~D. Pierro, and M.~Zaknoon, ``Steered sequential projections
  for the inconsistent convex feasibility problem,'' \emph{Nonlinear analysis:
  theory, method, and application, series A}, vol.~59, pp. 385--405, 2004.

\end{thebibliography}

\end{document}